\documentclass[a4paper]{article}

\usepackage{amsmath,amssymb}

\usepackage{anyfontsize}

\usepackage[utf8]{inputenc}
\usepackage[T1]{fontenc}

\usepackage[sfdefault]{biolinum}
\usepackage[sansmath]{libertinust1math}

\usepackage{hyperref}
\usepackage{cleveref}
\hypersetup{
  colorlinks=true,
  linkcolor=blue,
  filecolor=magenta,      
  urlcolor=cyan
}

\allowdisplaybreaks

\usepackage[a4paper,margin=1.3in]{geometry}

\usepackage{amsfonts}
\usepackage{graphicx}
\usepackage{algorithmic}
\usepackage{multibib}

\usepackage{amsmath}
\usepackage{amsthm}
\usepackage{amssymb}
\usepackage{braket}
\usepackage{url}
\usepackage{mathdots}
\usepackage[export]{adjustbox}
\usepackage{booktabs}
\usepackage{caption}
\usepackage{subcaption}
\usepackage{color} 
\usepackage[vlined, ruled]{algorithm2e}
\usepackage{bm}
\usepackage{diagbox}
\usepackage{multirow}
\usepackage{makecell}
\usepackage{rotating}

\newcommand{\llbracket}{{[\![}}
\newcommand{\rrbracket}{{]\!]}}

\newcommand{\floor}[1]{\lfloor{#1}\rfloor}

\usepackage[english]{babel}

\newcommand{\algoone}{\cite[Algo.~1]{patel2008optimal} }

\SetKwProg{Fn}{Function}{}{}

\title{Gaussian elimination versus greedy methods for the synthesis of
  linear reversible circuits\footnote{This document is the author's
    version of the corresponding research manuscript prior to formal
    peer review. An updated version is published in ACM Transactions
    on Quantum Computing, Volume 2, Issue 3, Article 11. pp 1–26,
    2021, \href{https://doi.org/10.1145/3474226}{10.1145/3474226}}}

\author{
Timothée Goubault de Brugière$^{1,3}$,
Marc Baboulin$^{1}$, Benoît Valiron$^{2}$,\\
Simon Martiel$^{3}$ and Cyril Allouche$^{3}$
}
\date{%
  $^{1}${\textit{Laboratoire de Recherche en Informatique, Université Paris-Saclay}, 
Orsay, France}\\
$^2${\textit{Laboratoire de Recherche en Informatique, CentraleSupélec}, 
Orsay, France}\\
$^3${\textit{Atos Quantum Lab}, Les Clayes-sous-Bois, France}}

\begin{document}

\maketitle

\begin{abstract}
  Linear reversible circuits represent a subclass of reversible
  circuits with many applications in quantum computing. These circuits
  can be efficiently simulated by classical computers and their size
  is polynomially bounded by the number of qubits, making them a good
  candidate to deploy efficient methods to reduce computational costs.
  We propose a new algorithm for synthesizing any linear reversible
  operator by using an optimized version of the Gaussian elimination
  algorithm coupled with a tuned LU factorization. We also improve the
  scalability of purely greedy methods.  Overall, on random operators,
  our algorithms improve the state-of-the-art methods for specific
  ranges of problem sizes: the custom Gaussian elimination algorithm
  provides the best results for large problem sizes ($n > 150$) while
  the purely greedy methods provide quasi optimal results when
  $n < 30$. On a benchmark of reversible functions, we manage to
  significantly reduce the CNOT count and the depth of the circuit
  while keeping other metrics of importance (T-count, T-depth) as low
  as possible.
\end{abstract}

\section{Introduction}

Progress in scientific computing in recent decades is as much about
improvements in algorithms as the increasing power of hardware. Some
very active fields, such as machine learning, have benefited from the
considerable increase in the amount of data available and from the
ever more efficient hardware capable of processing this data, whereas
the first algorithms were invented well before. This exponential
increase in the power of our computers is often illustrated, now for
40 years, by Moore's second law \cite{moore1975progress}, that
predicts that the number of transistors in microprocessors doubles
every 2 years since its formulation in 1975. Behind this law hides a
race for the miniaturization of transistors. Today the transistors are
only a few nanometers large, at this scale quantum effects such as the
tunnel effect interfere with the proper performance of the transistors
so that it becomes more and more complex to reduce their size.

New kind of computational models are developing in response to this
future limitation of the so-called ``conventional'' or ``classical''
computing. We can mention biological
\cite{guet2002combinatorial,gander2017digital}, photonic
\cite{kok2007linear}, neuromorphic \cite{monroe2014neuromorphic,
  zhao2010nanotube} or quantum \cite{nielsen2011quantum} computing,
booming these recent years. These new paradigms are promising because
they perform the calculation at new scales and potentially overcome
the limitations of classical computing. For example, in quantum
computing, elementary particles are directly manipulated by exploiting
their quantum properties to perform in polynomial time computations
that may require an exponential amount of operations on a conventional
machine \cite{shor1999polynomial,harrow2009quantum}.

Our paper concerns reversible computing, i.e., boolean operations that
can be ``undone'' because they consist in a bijection between the
inputs and outputs of the system. Reversible computing is a subclass
of photonic and quantum computing, which motivates its study.  A
consequence of the reversibility is that there is no loss of
information during the calculation, resulting in lower energy
consumption because no energy is dissipated at the erasure of a bit
\cite{landauer1961irreversibility}.  Here we study reversible circuits
referred to as linear reversible circuits. These circuits have the
particularity to be expressed as circuits containing only one
reversible logic gate: the Controlled-Not (CNOT) gate which performs
the operation NOT on a target qubit conditioned by the value of a
control qubit. This class of circuits is particularly useful in
quantum computing in the synthesis of stabilizer circuits: canonical
forms for stabilizer circuits involving linear reversible circuits
have been found and the size of stabilizer circuits is dominated by
the size of reversible linear circuits
\cite{aaronson2004improved,DBLP:journals/tit/MaslovR18} .  Stabilizer
circuits are for instance used for error correction and are therefore
essential for scaling up quantum computing
\cite{gottesman1997stabilizer}, and so are linear reversible circuits
by extension. Linear reversible circuits are also involved in circuits
called ``CNOT + T'' (also called phase polynomials) whose optimization
is the purpose of recent research \cite{DBLP:journals/tcad/AmyMM14,
  meuli2018sat, amy2018controlled, 2058-9565-4-1-015004}.

In the following we study the synthesis and optimization of linear
reversible circuits. We are given a linear reversible operator as a
Boolean matrix and we seek a quantum circuit that implements it as a
series of elementary operations.
Our contributions are the following :
\begin{itemize}
\item We propose GreedyGE: an optimized version of the Gaussian
  elimination algorithm for synthesizing any linear reversible
  operator. When we zero out the subdiagonal entries of the matrix,
  like in a standard Gaussian elimination, we perform the operations
  that minimize an ad-hoc cost function, resulting in less operations
  for the synthesis.  We study the theoretical worst case behavior of
  our algorithm. Overall, our algorithm is asymptotically optimal with
  theoretical and practical improvements over the state-of-the-art
  methods \algoone and the syndrome decoding based method
  \cite{DBLP:conf/rc/BrugiereBVMA20}.  In terms of computational time
  our algorithm is faster than the standard Gaussian elimination
  algorithm.
\item We propose a new way to compute the LU decomposition of a linear
  reversible operator. This provides us triangular operators $L$ and
  $U$ whose synthesis yield shorter circuits, especially when the
  circuit is expected to be "small enough". We will detail this notion
  of "small enough" later in the article. This optimization finds
  application in other synthesis methods relying on an LU
  decomposition, for instance the recent method we designed using the
  syndrome decoding problem \cite{DBLP:conf/rc/BrugiereBVMA20}.
\item We show that cost minimization techniques are useful for finding
  circuits whose expected size is "small enough".  We briefly review
  the existing algorithm and we propose our own cost function and
  implementation that increase the range of validity of this kind of
  technique.
\item We provide benchmarks of our methods and compare them to
  state-of-the-art algorithms. First we test our algorithms on random
  circuits of various sizes. Overall, with GreedyGE, we can synthesize
  circuits with more than 150 qubits in a few seconds while giving
  circuits that are at least 10\% smaller in average than the current
  state of the art. With purely greedy methods, we report quasi
  optimal results when $n < 30$.  Then we include our methods into a
  more general Clifford+T quantum compiler --- the Tpar algorithm
  \cite{DBLP:journals/tcad/AmyMM14} --- and we show that our
  algorithms are also useful in practice: on a set of reversible
  functions we successfully reduce the total number of CNOTs and the
  circuit depth while keeping other metrics of interest (T-count,
  T-depth) as low as possible.
\end{itemize}

The plan of this paper is the following: in Section~\ref{background}
we present the basic notions and state of the art about the synthesis
of linear reversible circuits. In Section~\ref{cnot_size::greedyge} we
describe the improved Gaussian elimination algorithm on triangular
operators. In Section~\ref{cnot_size::extension} we extend our
algorithm to treat any general linear reversible operator. Notably we
present optimization techniques for computing the LU decomposition. We
tackle the study of purely greedy methods in
Section~\ref{cnot_size::pathfinding}. Benchmarks are given in
Section~\ref{cnot_size::bench}.  We conclude in
Section~\ref{cnot_size::conclusion}.

\section{Background and state of the art} \label{background}

\subsection{Notion of linear reversible function.}
Let $\mathbb{F}_2$ be the Galois field of two elements. A
boolean function $f : \mathbb{F}_2^n \to \mathbb{F}_2$ is
said to be linear if
\[
  f(x_1 \oplus x_2) = f(x_1) \oplus f(x_2)
\]
for any $x_1, x_2 \in \mathbb{F}_2^n$ where $\oplus$ is the bitwise
XOR operation. Let $e_k$ be the $k$-th canonical vector of
$\mathbb{F}_2^n$. By linearity we can write for any
$x = \sum_k \alpha_k e_k$ (with $\alpha_k \in \{0,1\}$)
\[
  f(x) = f\left(\sum_{k} \alpha_k e_k\right) = \sum_k \alpha_k
  f(e_k)
\]
and the function $f$ can be represented with a column vector
$\mathbf{c} = [f(e_1), ..., f(e_n)]^T$ such that $f(x) = c^T x$.  This
expression easily extends to the $n$-inputs and $m$-outputs functions
$f : \mathbb{F}_2^n \to \mathbb{F}_2^m$ where $f$ is defined by an
$m \times n$ boolean matrix $A$ such that
\[ f(x) = Ax. \]

In the case of reversible boolean functions we have $n=m$ and the
matrix $A$ must be invertible. The application of two successive
operators $A$ and $B$ is equivalent to the application of the operator
product $BA$. There is a one-to-one correspondence between the linear
reversible functions of arity $n$ and the invertible boolean matrices
of size $n$. This was used for instance to count the number of
different linear reversible functions of $n$ inputs in
\cite{patel2008optimal}.

\subsection{LU decomposition.}
Given the matrix representation $A$ of a generic linear reversible
operator, we can always perform an LU decomposition~\cite{GVL96} such
that there exists an upper (resp. lower) triangular matrix $U$
(resp. $L$) and a permutation matrix $P$ such that $A = PLU$.  The
invertibility of $A$ ensures that the diagonal elements of $L$ and $U$
are all equal to $1$. In the remainder of this paper, the term
``triangular operator'' stands for an operator whose corresponding
matrix is either upper or lower triangular. The LU decomposition is at
the core of several synthesis of general linear reversible Boolean
operators: synthesizing $U$, $L$, $P$ and concatenating the circuits
gives an implementation of $A$.

\subsection{Synthesis of linear reversible boolean functions.}
We are interested in synthesizing a linear reversible boolean
function into a reversible circuit, i.e., a series of elementary
reversible gates that can be executed on suitable hardware. For
instance in quantum computing the CNOT gate is used
for superconducting and photonic qubits and
performs the following 2-arity operation:
\[
  \text{CNOT}(x_1, x_2) = (x_1, x_1 \oplus x_2).
\]
The CNOT gate is a linear reversible gate. It is also universal for
linear reversible circuit synthesis, i.e., any linear reversible
function can be implemented by a reversible circuit containing only
CNOT gates \cite{aaronson2004improved}.  In this paper we aim at
producing CNOT-based reversible circuits for any linear reversible
functions.

In terms of matrices, a CNOT gate controlled by the line $j$ acting on
line $i \neq j$ can be written $E_{ij} = I + e_{ij}$ where $I$ is the
identity matrix and $e_{ij}$ the elementary matrix with all entries
equal 0 but the entry $(i,j)$ whose value is $1$.  Finding a CNOT
circuit implementing an operator $A$ is therefore equivalent to
finding a sequence of matrices $(E_{i_kj_k})_{1\leq i_k,j_k \leq n}$
such that
\[ \prod_{k=1\dots N} E_{i_k,j_k} = A. \] Using the fact that
$E_{ij}^{-1} = E_{ij}$, it is more convenient to rewrite the synthesis
problem as a reduction of $A$ to the identity operator $I$
\[ \prod_{k=N\dots 1} E_{i_k,j_k}A = I\] because we now show that the
synthesis problem can be reformulated in terms of elementary
operations applied on $A$.  Given a CNOT with control $j$ and target
$i$ applied on an operator $A$, the updated operator $E_{ij}A$ can be
deduced from $A$ with an elementary row operation:
\[
  r_i \leftarrow r_i \oplus r_j,
\]
writing $r_k$ for the $k$-th row of $A$. Therefore, if one can compute
a sequence of row operations transforming $A$ into the identity
operator then one can construct a circuit implementing $A$ by
concatenating the CNOT gates associated to each row operation. With a
standard Gaussian elimination algorithm it is always possible to
reduce an invertible operator to the identity with row
operations. This gives a simple proof that the CNOT gate is universal
for linear reversible circuit synthesis. In some cases we will also
authorize the use of column operations. A column operation
$c_j \leftarrow c_i \oplus c_j$ is equivalent to right multiplying $A$
by $E_{ij}$. Overall by combining row and column operations we get
\[
  \prod_{(i_1,j_1)} E_{i_1j_1} \times A \times
  \prod_{(i_2,j_2)} E_{i_2j_2} = I
\]
and finally 
\[
  A = \prod_{(i_1,j_1)} E_{i_1j_1} \prod_{(i_2,j_2)} E_{i2j2} 
\]
so we recover a CNOT circuit for the implementation of $A$.

Thus, synthesizing a linear reversible function into a CNOT-based
reversible circuit is equivalent to transforming an invertible Boolean
matrix $A$ to the identity by applying elementary row and column
operations.  From now on we will privilege this more abstract point of
view because it gives more freedom and often appears clearer for the
design of algorithms. We note by $\text{Row}(i,j)$ the elementary row
operation $r_j \leftarrow r_i \oplus r_j$ and $\text{Col}(i,j)$ the
elementary column operation $c_j \leftarrow c_i \oplus c_j$.

We use the size of the circuit, i.e., the number of CNOT gates in it,
to evaluate the quality of our synthesis. The size of the circuit
gives the total number of instructions the hardware has to perform
during its execution. Due to the presence of noise when executing
every logical gate, it is of interest to have the shortest circuit
possible. Another metric of interest is the depth of the circuit which
is closely related to the number of time steps required to execute the
circuit. In other words the depth of a quantum circuit gives its
execution time. Given that the physical qubits are subject to the
decoherence problem and that the total time available to perform a
quantum algorithm is limited, it is also very important to be able to
produce shallow circuits.

We now review the main algorithms designed for the synthesis of CNOT circuits.

\subsection{State of the art and contributions}

We distinguish algorithms that assume that all qubits are
connected and algorithms that take into account a restricted
connectivity between the qubits. The former usually give smaller and
shallower circuits while the latter are adapted to real world hardware
where row operations can be performed only between qubits that are
neighbors in the hardware. Studying ideal cases with full connectivity
is of interest for providing us with new solutions before transposing
them to the restricted case.

\paragraph{Gaussian elimination based algorithm.} In the literature,
the first algorithm proposed is a Gaussian elimination algorithm that
produces circuits of size $\mathcal{O}(n^2)$ gates
\cite{aaronson2004improved}. Later a lower bound of
$\mathcal{O}(n^2/\log_2(n))$ was established in
\cite{patel2008optimal} for the asymptotic optimal size of reversible
circuits, with an algorithm that reaches this lower bound. The
algorithm consists in performing the Gaussian elimination algorithm on
a chunk of $m$ columns. The gain in the number of operations arises
when two words in $F_2^m$ are equal and that one can be zeroed in one
row operation. Choosing $m = \mathcal{O}(\log_2(n))$ leads to an
optimal asymptotic complexity.

\paragraph{Syndrome decoding based algorithm} Recently, we proposed a
new method for the synthesis of CNOT circuits
\cite{DBLP:conf/rc/BrugiereBVMA20}. We transformed the synthesis
problem into a series of syndrome decoding problems and gave several
strategies to solve them (either optimal with an integer programming
solver or approximate with greedy methods). Our benchmarks reveal that
our method outperforms \algoone for operators on medium-sized
registers ($n < 400$ qubits) but performs poorly as $n$ gets larger.

\paragraph{Other algorithms.} In \cite{schaeffer2014cost} a
cost-minimization approach was used with promising results.

\bigskip To our knowledge, \algoone is state-of-the-art and it is used
in the Tpar and Gray-Synth algorithms
\cite{amy2018controlled,DBLP:journals/tcad/AmyMM14}. Given that our
recent results from \cite{DBLP:conf/rc/BrugiereBVMA20} improve
\algoone, we also have to take them into consideration. Therefore,
both \algoone and the syndrome decoding based algorithms will be our
state-of-the-art methods in this paper.

\paragraph{Related works.} Recently an algorithm asymptotically tight
for the depth was also proposed
\cite{DBLP:conf/soda/JiangSTW0Z20}. All the algorithms we mentioned so
far assume that all qubits are connected. For special hardware
constraints, some variants have been designed. In the case of a Linear
Nearest Neighbor (LNN) architecture, the works in
\cite{maslov2007linear,DBLP:journals/cjtcs/KutinMS07} provide circuits
with linear depth. For a generic qubit connectivity, the works in
\cite{DBLP:journals/qic/KissingerG20,nash2020quantum} propose an
adaptation of the Gaussian elimination method using Steiner trees and
reach circuits of size $\mathcal{O}(n^2)$. The syndrome decoding based
algorithm has also been extended to arbitrary connectivities
containing a Hamiltonian path \cite{DBLP:conf/rc/BrugiereBVMA20}.

\paragraph{Contributions.} In this paper we focus on improving the
size of the generated circuits when considering a full qubit
connectivity. To our knowledge {\algoone} and the algorithm from
\cite{DBLP:conf/rc/BrugiereBVMA20} are in this case the best
algorithms. The former is polynomial in time and produces
asymptotically optimal circuits size. The latter is also polynomial in
time if the algorithm for solving the syndrome decoding problem has a
polynomial complexity. Better results can be obtained with an exact
--- but exponentially costly --- solver but, in any case, the syndrome
decoding based algorithm produces shorter circuits for medium sized
circuits, without any theoretical guarantee though.

Our method relies on an efficient algorithm --- GreedyGE --- for
synthesizing any triangular operator: at any time during the synthesis
the next row operation is chosen such that it minimizes a custom cost
function. Then we propose two extensions to be able to synthesize any
generic operator. The first option is similar to \algoone: we slightly
modify our algorithm such that it first transforms $A$ into an upper
triangular operator $U$ and then we use our unmodified algorithm to
synthesize $U$.  The second option is to rely on the LU decomposition
of the operator to synthesize: given $A = PLU$, the synthesis of $A$
can be performed by synthesizing $U$, $L$ and $P$ and concatenating
the circuits.  The permutation $P$ can be synthesized by applying
successive SWAP gates that requires 3 CNOTs each. Yet, one can avoid
applying the permutation by doing a post-processing of the circuit
that would transfer the permutation operation directly at the end of
the total circuit. In the case of a full connectivity this can be done
without any overhead in the gate count. Consequently for that second
option we do not consider the cost of implementing the permutation $P$
and the cost of synthesizing $A$ is twice the cost of synthesizing a
triangular operator.

Our algorithm improves \algoone and \cite{DBLP:conf/rc/BrugiereBVMA20}
in several ways :
\begin{itemize}
\item first we improve the worst case synthesis of triangular
  operators with theoretical guarantee. Namely we show that the size
  of our circuits is upper bounded by $n^2/\log_2(n)$ and that we
  converge faster than \algoone to that asymptotic behavior.
\item We propose a fast implementation of GreedyGE such that the
  running time is lower than the standard Gaussian elimination
  algorithm.
\item We propose a new way to compute LU decompositions such that we
  get triangular operators whose synthesis leads to shorter
  circuits. This improvement can also be used in
  \cite{DBLP:conf/rc/BrugiereBVMA20} as it also relies on an LU
  decomposition.
\end{itemize}

Overall GreedyGE is suited to dense, worst case operators, or
operators that act on several hundred or thousands of qubits. For
operators on a much smaller number of qubits, or when the structure of
the operator is such that we can expect the resulting circuit to be
small (for instance if the operator is sparse), GreedyGE does not
particularly outperform other existing algorithms. Instead we show
that cost minimization techniques, like the one introduced in
\cite{schaeffer2014cost}, produce much better results. We review the
work in \cite{schaeffer2014cost} and we propose a new cost function
that improves the range of validity of such cost minimization method.

Finally we apply our new methods to real world circuits. Plugged into
a global Clifford+T quantum compiler, here the Tpar algorithm
\cite{DBLP:journals/tcad/AmyMM14}, we show that we can significantly
improve the CNOT cost of reversible functions without worsening other
metrics like the T-depth.

\section{GreedyGE: a Greedy Gaussian Elimination Algorithm
  for\texorpdfstring{\\*}{ }Triangular Boolean
  Matrices} \label{cnot_size::greedyge}

\subsection{General Presentation of the Algorithm}

Given a lower triangular operator $L$ --- the upper case can be
treated similarly --- the Gaussian elimination algorithm can be
summarized as follows: by applying elementary row operations
$\text{Row}(i,j)$ with $i < j$ we zero the sub-diagonal elements of
$L$ without changing its triangular shape. This process is performed
column by column, starting from the first one. This way one can easily
see that if the k first columns are treated, applying row operations
$\text{Row}(i,j)$ with $k < i < j$ will not change the treated columns
as all their elements are 0. Since $L$ is invertible, when zeroing the
$k$-th column the $k$-th row is necessarily equal to $L[k,:] = e_k^T$
hence one can always add the row $k$ to any row $j > k$ to zero the
entry $L[j,k]$. This guarantees the good behavior of the algorithm for
any input matrix.

Usually, a Gaussian elimination algorithm always performs the row
operations $\text{Row}(k,j)$ when zeroing the $k$-th column. As one
row operation ensures to zero only one element at a time, we need at
most $\frac{n(n-1)}{2}$ row operations to zero every sub-diagonal
elements of $L$, leading to a worst-case complexity of $O(n^2)$ for
the synthesis of a generic linear reversible operator.

A first straightforward improvement is to authorize any row operation
$\text{Row}(i,j)$, $i<j$. This was exploited in \algoone: by
partitioning the matrix of size $n$ into blocks of size $m$, the
authors use the fact that there can only be at most $2^m$ different
vectors in each block. So for $n > 2^m$ it is possible to zero $n-2^m$
full vectors of size $m$ by applying one row operation for each
vector. Using this method some row operations can zero up to $m$
entries at a time, diminishing the total number of row operations. By
choosing $m = \floor{\alpha \log_2(n)}$ (for arbitrary $0<\alpha<1$)
they reach an asymptotic complexity of $n^2/(\alpha\log_2(n))$ in the
worst case which meets the theoretical lower bound.

Our proposed algorithm can be seen as a direct improvement of
\algoone. It comes from the following simple observation: given a row
$i$ of our triangular operator and assuming that we follow a standard
Gaussian elimination process, an upper bound on the number of row
operations that will be applied to row $i$ during the synthesis is
given by
\begin{equation} 
\#\text{CNOTs} = i - \min\{j \; | \; L[i,j] = 1 \}.
\label{greedy}
\end{equation}
In other words, when zeroing the entries of a row, we have to focus
first on the entries on the first columns and this can be done by
minimizing the cost function given by Eq.~\eqref{greedy}. Once an
entry is zeroed, we are sure that it is ``treated'' and will never be
modified.  Note on the other hand how zeroing entries in the last
columns do not give us the guarantee that they would not be modified
again during the synthesis.  This simple observation is at the core of
the Gaussian elimination algorithm: entries are acted upon one by
one. It is also at the core of {\algoone} which improves on regular
Gaussian elimination: entries are acted upon by fixed-size blocks. In
this paper, we improve on {\algoone} by allowing for varying-size
blocks.  In other words, we do not fix a block size but at each
iteration we let the greedy component of our algorithm choose the
largest suitable block size.

Therefore we propose the following 3-step method:
\begin{enumerate} 
\item choose among the rows with entries "1" in the left-most columns
  the two rows $i$ and $j$ that have the largest number of common
  left-non-zero-elements,
\item apply the row operation $\text{Row}(\min(i,j), \max(i,j))$,
\item repeat the first two steps until $L$ is the identity.
\end{enumerate}

The pseudo-code of the algorithm is given in Algorithm
\ref{PseudoCode}. The time complexity of the algorithm mostly depends
on step 1, i.e., the function SelectRowOperation. Actually, this step
can be easily implemented iteratively on the columns :
\begin{itemize}
\item We start with the set of all the rows and, during the $k$-th
  step, we have in memory a set of rows having the same first $k-1$
  entries.
\item Then we separate those rows into two sets: those which have the
  $k$-th entry respectively equal to $0$ and $1$, corresponding
  respectively to the sets $\text{set}_0$ and $\text{set}_1$ in
  Algorithm~\ref{PseudoCode}. The rows in either $\text{set}_0$ or
  $\text{set}_1$ have the same first $k$ entries.
\item We want to zero the maximum of $1$ in priority so if
  $\text{set}_1$ contains $2$ or more rows we continue with this set,
  otherwise we choose $\text{set}_0$ and go to step k+1.
\end{itemize}
By this process, the size of our set of lines always remains greater
than $2$ until we have only two lines: these are the two lines on
which to do our row operation. Clearly, the algorithm is polynomial in
the size of the matrix.

\begin{algorithm}
\caption{GreedyGE : Greedy Gaussian Elimination of a triangular operator $L$.}
\label{PseudoCode}
\begin{algorithmic} 
\REQUIRE $n \geq 0, \; \; L \in \mathbb{F}_2^{n \times n}$ lower triangular
\ENSURE $C$ is a CNOT-circuit implementing $L$
\STATE
\Fn{Synthesis(L, n)}
{
\STATE $C \leftarrow [\;]$
\WHILE{$L \neq I_n$}
\STATE set $\leftarrow \text{SelectRowOperation}(L,n)$
\STATE $i_0 \leftarrow \min(\text{set}[1], \text{set}[2])$
\STATE $j_0 \leftarrow \max(\text{set}[1], \text{set}[2])$
\STATE C.append(CNOT($i_0,j_0$))
\STATE $L[j_0,:] \leftarrow L[i_0,:] \oplus L[j_0,:]$
\ENDWHILE
\RETURN reverse(C)
}
\STATE
\Fn{SelectRowOperation(L, n)}
{
\STATE set $\leftarrow \llbracket 1, n \rrbracket$
\STATE $j \leftarrow 1$
\WHILE{$|\text{set}| > 2$}
\STATE $\text{set}_0 \leftarrow \{ i \in \text{set} \; | \; L[i,j] = 0 \}$
\STATE $\text{set}_1 \leftarrow \{ i \in \text{set} \; | \; L[i,j] = 1 \}$
\IF{|$\text{set}_1$| < 2}
\STATE set $\leftarrow$ $\text{set}_0$
\ELSE 
\STATE set $\leftarrow$ $\text{set}_1$
\ENDIF
\STATE $j \leftarrow j+1$ 
\ENDWHILE
\RETURN set
}
\end{algorithmic}
\end{algorithm}

Our algorithm belongs to the same family as the Gaussian elimination
method and \algoone because we synthesize the operator column by
column. However, this is an improvement over \algoone because we do
not work with a fixed block size. At each step, we somewhat choose the
largest block size possible to zero the largest number of elements in
one row operation. We are then ensured that these entries will not be
modified anymore. Thus this is also an improvement over cost
minimization methods \cite{schaeffer2014cost} that can take as cost
function the number of ones in the matrix for instance. With such
greedy methods, it may be possible that in one step more elements are
zeroed but it is not guaranteed that theses elements will be left
untouched thereafter. We keep a structure in the synthesis which
enforces the convergence of our greedy algorithm.

\subsection{Improving the Time Complexity}

When executing Algorithm~\ref{PseudoCode} we find at each iteration
the next row operation by selecting the rows according to their number
of common first elements. As a first approach, we keep the set of the
most promising rows until the size of the set is equal to 2. With a
little more work we can in fact order all the rows in one run and get
the whole set of row operations required to zero the sub-diagonal
elements of the current column. It is more efficient than redoing the
sorting work after each new row operation because the selection
algorithm will repeat the same calculations again and again but
without the presence of the modified rows. The new pseudo-code for a
more efficient version of our method is given
Algorithm~\ref{FasterPseudoCode}.

\begin{algorithm}
\caption{FastGreedyGE : Greedy Gaussian Elimination of a triangular operator $L$.}
\label{FasterPseudoCode}
\begin{algorithmic} 
\REQUIRE $n \geq 0, \; \; L \in \mathbb{F}_2^{n \times n}$ lower triangular
\ENSURE $C$ is a CNOT-circuit implementing $L$
\STATE
\Fn{OptimizedSynthesis(L, n)}
{
\STATE $C \leftarrow [\;]$
\FOR{j = 1 to n-1}
\STATE set $\leftarrow \{ i \in [j,...,n] \; | \; L[i,j] = 1 \}$
\STATE pairs, set = SelectAllRowOperations(L, j+1, set)
\FOR{pair = pairs}
\STATE $i_0 \leftarrow \min(\text{pair}[1], \text{pair}[2])$
\STATE $j_0 \leftarrow \max(\text{pair}[1], \text{pair}[2])$
\STATE C.append(CNOT($i_0,j_0$))
\STATE $L[j_0,:] \leftarrow L[i_0,:] \oplus L[j_0,:]$
\ENDFOR
\ENDFOR
\RETURN reverse(C)
}

\Fn{SelectAllRowOperations(L, j, set)}
{
\IF{|set| < 2}
\RETURN $\emptyset$, set
\ENDIF
\STATE $\text{set}_0 \leftarrow \{ i \in \text{set} \; | \; L[i,j] = 0 \}$
\STATE $\text{set}_1 \leftarrow \{ i \in \text{set} \; | \; L[i,j] = 1 \}$
\STATE $\text{pairs}_0, \text{set}_0 \leftarrow \text{SelectAllRowOperations}(L, j+1, \text{set}_0)$
\STATE $\text{pairs}_1, \text{set}_1 \leftarrow \text{SelectAllRowOperations}(L, j+1, \text{set}_1)$

\STATE pairs $\leftarrow [\text{pairs}_0, \text{pairs}_1]$
\IF{$|\text{set}_0| > 0 \; \AND \; |\text{set}_1| > 0$}
\STATE $i_0 \leftarrow \text{set}_0[1]$
\STATE $i_1 \leftarrow \text{set}_1[1]$
\STATE pairs $\leftarrow [\text{pairs}, [\min(i_0,i_1), \max(i_0,i_1)]]$
\STATE set $\leftarrow \min(i_0, i_1)$
\ELSE
\IF{$|\text{set}_0| > 0$}
\STATE set $\leftarrow \text{set}_0[1]$
\ELSE
\STATE set $\leftarrow \text{set}_1[1]$
\ENDIF
\ENDIF
\RETURN pairs, set
}
\end{algorithmic}
\end{algorithm}

\begin{figure*}[p!]
\begin{center} \[L = \begin{pmatrix} 1  &0  &0 & 0 & 0 & 0 & 0 & 0 & 0 & 0  \\
1 & 1 & 0 & 0 & 0 & 0 & 0 & 0 & 0 & 0  \\
1 & 1 & 1 & 0 & 0 & 0 & 0 & 0 & 0 & 0  \\
1 & 1 & 1 & 1 & 0 & 0 & 0 & 0 & 0 & 0  \\
0 & 1 & 0 & 0 & 1 & 0 & 0 & 0 & 0 & 0  \\
1 & 1 & 0 & 0 & 0 & 1 & 0 & 0 & 0 & 0  \\
1 & 1 & 1 & 0 & 0 & 1 & 1 & 0 & 0 & 0  \\
1 & 0 & 1 & 1 & 1 & 0 & 0 & 1 & 0 & 0  \\
1 & 0 & 0 & 0 & 0 & 1 & 0 & 1 & 1 & 0  \\
0 & 0 & 0 & 1 & 0 & 0  &1 & 1 & 1 & 1 \end{pmatrix} \] 

\hspace*{-1.5cm}\includegraphics[scale=0.6]{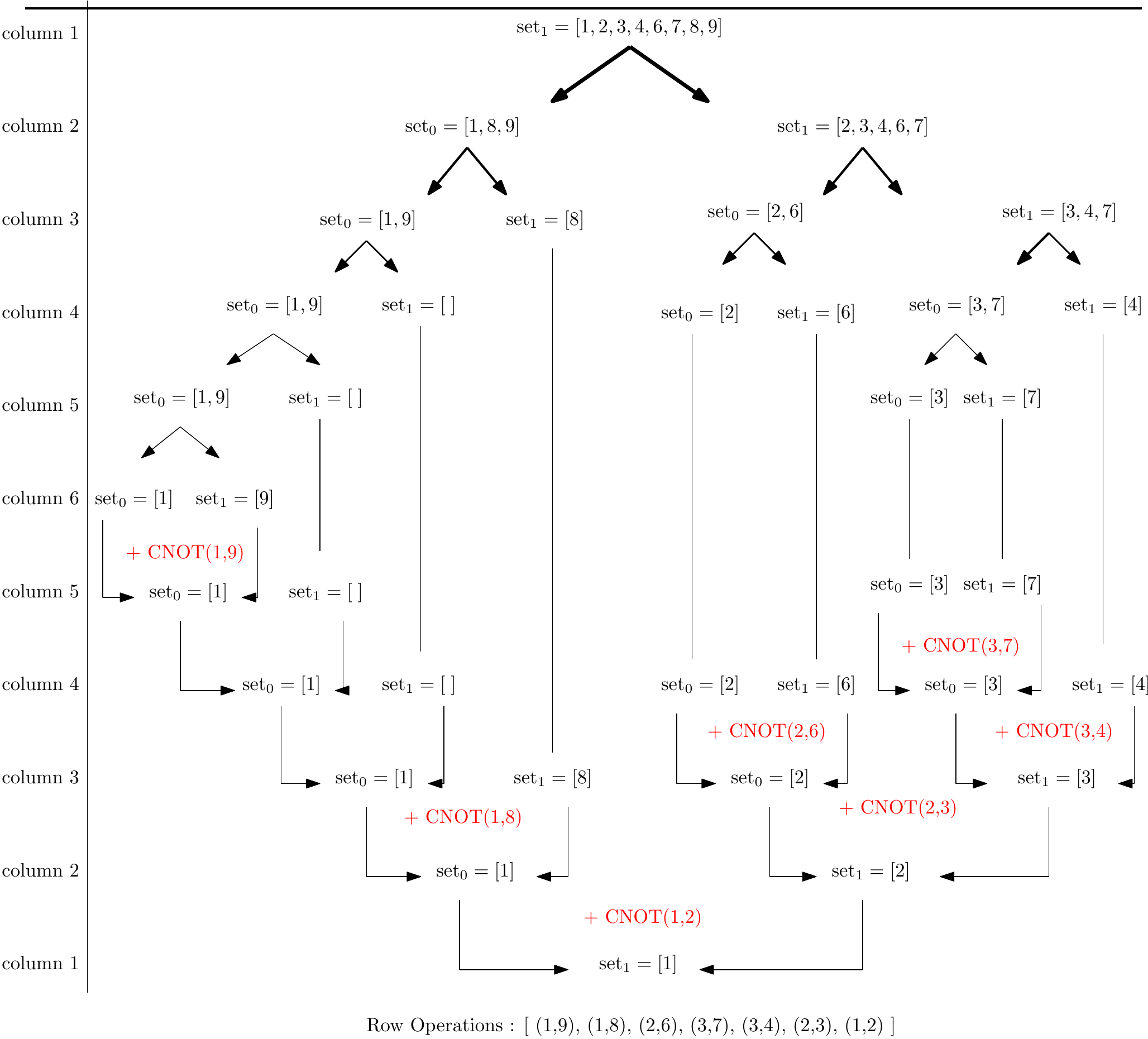}
\end{center}
\def\mycap{Illustration of the SelectAllRowOperations function on a specific example.}
\caption{\mycap}
\label{example}
\end{figure*}

The new version of SelectRowOperation, the function
SelectAllRowOperations, is based on the idea that row operations
should always be done, if possible, between rows belonging to the same
sets. We implement this idea recursively on the columns: first, we
only select the rows with a "1" entry in the first column. Then, after
the creation of set0 and set1, we call recursively
SelectAllRowOperations on both sets: we have two sets of row
operations that treat separately the rows with first entries "10" and
"11". After the execution of SelectAllRowOperations, both set0 and
set1 contains the unmodified rows. In both cases, there can be at most
one row unmodified. If both sets have one row unmodified then we add
another row operation
$\text{Row}(\min(\text{set}_0[1], \text{set}_1[1]),
\max(\text{set}_0[1], \text{set}_1[1]))$ and the set returned contains
$\min(\text{set}_0[1], \text{set}_1[1])$. If only one set has an
unmodified row then the set returned contains that row. We illustrate
the behavior of our algorithm with the example given in
Fig~\ref{example}.

\subsection{Bounding the CNOT Count}

We denote with $C(n)$ the number of CNOTs required to synthesize a
triangular operator $L$ of size $n$ with Algorithm
\ref{PseudoCode}. Given $1 \leq k \leq n$, we consider $c(n,k)$ the
number of row operations required to synthesize the first $k$ columns.

Once the first $k$ columns are synthesized the $n \times k$ block
matrix $L[:,1:k]$ is equal to the matrix
$\begin{pmatrix} I_{k} \\ 0 \end{pmatrix}$ and we are left with the
sub-matrix $L[k+1:end,k+1:end]$ to synthesize. Then we have
\[
  \forall k \geq 1, \; C(n) \leq c(n,k) + C(n-k)
\]
and the total number of row operations to synthesize $L$ is upper
bounded by
\begin{equation}
  C(n) \qquad \leq \qquad \sum_{i=1}^m c \left(n - \sum_{j=1}^{i-1} k_j, \, k_i\right)
  \label{upper_bound}
\end{equation}
where $k_1, k_2, ..., k_m$ can represent any partition of $L$ into $m$
sub-blocks, i.e., $m$ positive integers such that
$\sum_{i=1}^m k_i = n$. By considering the worst-case scenario for
each block we do not take into consideration the side effects of the
row operations that were applied on the first blocks and that could
have an impact on the remaining matrix. Especially when looking
strictly at one block, one can easily see that some row operations
will only zero one element at a time in a block --- the rightmost
elements --- but will certainly have an impact on the next block: by
not considering this effect, we weaken our upper bound but the proof
of the inequality \eqref{upper_bound} is straightforward.

We now turn to an estimation of an upper bound of
$c(n,k)$. Considering a block of size $(n,k)$ and assuming that
$k < \log_2(n)$, the algorithm process is the following :
\begin{itemize}
\item as there can be at most $2^k$ different bitstrings of $k$ bits,
  $n-2^k$ row operations at most will be used to zero the duplicate
  rows.
\item We are left with a rectangular block of size $(2^k,k)$. Using a
  similar argument we can zero $k-1$ bits of $2^{k-1}$ different rows
  by using one row operation for each, then the remaining bit on these
  rows can be zeroed with one row operation. We are now left with a
  block of size $(2^{k-1},k)$, we zero $k-2$ bits of $2^{k-2}$ rows
  with one row operation each and the two remaining bits on these rows
  can be zeroed with two row operations. Repeating this process until
  the whole block has been reduced to
  $\begin{pmatrix} I \\ 0 \end{pmatrix}$, we perform at most
  \begin{equation*} 2^{k-1} \times (1+1) + 2^{k-2} \times (1+2) +
    2^{k-3} \times (1+3) ... = 2^{k+1} \times \sum_{j=2}^{k+1} j/2^j
    \leq 3 \times 2^k\end{equation*} row operations. We can save some
  row operations because the right upper triangular part of the block
  is already zeroed (as $L$ is triangular) but we neglect them for
  simplicity.
\end{itemize}

So we have the estimation
\begin{equation}
  c(n,k) \leq n + 2^{k+1}.\label{upper_bound_block}
\end{equation}

From this we can derive an upper bound for $C(n)$. We replace
Eq.~\eqref{upper_bound_block} in Eq.~\eqref{upper_bound} :
\[ C(n) \leq \sum_{i=1}^m \left(n - \sum_{j=1}^{i-1} k_j + 2^{k_i+1}
  \right).\] As $n - \sum_{j=1}^{i-1} k_j = \sum_{j=i}^m k_j$ for any
$i$ we rewrite the upper bound
\begin{equation*}
  C(n) \leq \sum_{i=1}^m \left(2^{k_i+1} + \sum_{j=i}^m k_j\right)
\end{equation*}
\begin{equation}
  C(n) \leq \sum_{i=1}^m \left(2^{k_i+1} + ik_i
  \right). \label{tight_upper_bound}
\end{equation}

For simplicity we assume that $k_1 = k_2 = ... = k_{m-1} = k$. We also
assume that $k_m = n - \sum_{i=1}^{m-1} k_i = k$ and $m = n/k$ because
it does not affect the results.

Replacing in Eq.~\eqref{tight_upper_bound}:
\begin{equation} C(n) \leq m2^{k+1} + k \frac{m(m+1)}{2} =
  \frac{n^2}{2k} + \frac{n}{k} 2^{k+1} +
  \frac{n}{2} \label{simplified_upper_bound}
\end{equation}

We now prove two things: 
\begin{enumerate}
\item we show that our upper bound is better than the one given in
  \cite{patel2008optimal}.
\item we prove that the asymptotic complexity is at most
  $n^2/\log_2(n)$.
\end{enumerate}

1. Taking $k = \alpha \log_2(n)$ with $0 < \alpha < 1$, we have
\begin{equation} 2 \times C(n) < \frac{n^2}{\alpha \log_2(n)} +
  \frac{4n^{1+\alpha}}{\alpha \log_2(n)} +
  n.\label{upper1} \end{equation} where the multiplication by $2$
gives an upper bound for the synthesis of a generic operator. This is
a direct improvement over the upper bound in \cite{patel2008optimal}
given by
\begin{equation} \text{total row op. \cite{patel2008optimal}} \leq
  \frac{n^2}{\alpha \log_2(n)} + 2n^{1+\alpha} + \text{ negligible
    terms}\label{markov_upper} \end{equation} provided
$\frac{4}{\alpha \log_2(n)} < 2$ which is the case if $n$ is
sufficiently large. For instance, with $\alpha = 1/2$ then $n > 16$ is
sufficient.

\medskip
\noindent
2. The coefficient of the leading term in Eq~\eqref{upper1} is
$1/\alpha$, yet taking the limit $\alpha \to 1$ in Eq.~\eqref{upper1}
leads to a complexity of $5n^2/\log_2(n)$. Instead, taking
$k = \log_2(n) - p$ for some $p$, we have
\[ 2 \times C(n) < \frac{n^2}{\log_2(n)} \times \frac{1}{1 -
    p/\log_2(n)} \times \left(1 + \frac{4}{2^p} \right) + n. \] By
choosing carefully $p$, for instance $p = \sqrt{\log_2(n)}$, we get
\[ 2 \times C(n) < a(n) \times \frac{n^2}{\log_2(n)} + n\] with
$a(n) = \frac{1}{1 - 1/\sqrt{\log_2(n)}} \times \left(1 +
  \frac{4}{2^{\sqrt{\log_2(n)}}}\right)$. As $a(n) \to 1$ this proves
that the asymptotic complexity is $n^2/\log_2(n)$.

We improve the upper bound Eq.~\eqref{markov_upper} in two ways: we
ensure that the coefficient of the leading term is $1$ and we avoid
other terms that distort the real complexity. Especially if one wants
to have the best asymptotic complexity with \algoone by choosing
$\alpha$ arbitrarily close to 1, the term $n^{1+\alpha}$ becomes
arbitrarily close to $n^2$ and its impact is not negligible until very
large values of $n$. Even with intermediate values of $n$, for example
$n=1000$, then $a(n) \approx 2.12$. To have the same leading term we
have to choose $\alpha = 1/2.12$. With these values we now have
\[ \frac{2n^{1+\alpha} + n^2/(\alpha\log_2(n))}{n^2/(\alpha\log_2(n))} \approx 1.25\]
which means that the term $2n^{1+\alpha}$ increases the upper bound by 25\%. 

\paragraph{Discussion.}

The main result we want to emphasize is that the upper bound in
Eq.~\eqref{tight_upper_bound} is true for any partitioning of the
matrix $L$. This means it performs better than any partitioning method
unless a more efficient method for synthesizing one block is
found. Notably, given $k_1 = k_2 = ... = k_{m-1} = k$ the size of the
block used for \algoone, it is clear that the row operations done in
\algoone to zero an entire k-bit row will also be performed by our
algorithm. Then we are left with a sub-matrix to synthesize for which
\algoone will use a Gaussian Elimination method whereas our algorithm
will perform recursively by first zeroing the bitstrings of size
$k-1$. In the worst case our algorithm will only zero one element at a
time like the Gaussian elimination algorithm.

This gives insights about the advantage of GreedyGE over
\algoone. When \algoone performs a standard Gaussian elimination
process to zero some elements, we somewhat perform \algoone on a
smaller sub-matrix. Thus, at these steps of the algorithm, we expect
to have the same gain in the number of row operations than \algoone
has compared to the classical Gaussian elimination
algorithm. Consequently, although we cannot theoretically claim that
our algorithm always generates shorter circuits than \algoone --- as
we cannot claim that \algoone always performs better than standard
Gaussian elimination --- we have strong confidence that our algorithm
provides better results most of the time --- for the same reasons that
\algoone gives most of the time better results than the Gaussian
elimination algorithm. This assertion is empirically verified in our
benchmarks, see Chapter~\ref{cnot_size::bench}.

\section{Extending GreedyGE to General Operators} \label{cnot_size::extension}

So far we have presented an algorithm for the synthesis of triangular
operators. To extend this to generic linear reversible operators $A$
we give two strategies:
\begin{itemize}
\item we can modify our algorithm for triangular operators such that
  it transforms $A$ into an upper triangular operator $U$ on which we
  apply our unmodified algorithm,
\item we can rely on an LU decomposition to factorize $A$ into the
  product of two triangular operators. Then the concatenation of the
  circuits implementing $U$ and $L$ yields a circuit implementing $A$.
\end{itemize}

\subsection{Modification of FastGreedyGE}

Algorithm~\ref{FasterPseudoCode} works even if the input operator is
not lower triangular. In the outer for-loop, during the $j$-th
iteration the algorithm will zero the sub-diagonal elements of the
$j$-th column provided that the element $A[j,j]$ is equal to $1$. So
all we need to do is adding an extra row operation to make sure that
$A[j,j]=1$ is satisfied. If $A[j,j]=1$ then we do nothing otherwise we
choose among the rows $\llbracket j+1, n \rrbracket$ whose $j$-th
element is $1$ the one that zeroes the largest number of right
elements of the j-th row so that when executing FastGreedyGE on the
future $U$ operator we save some row operations. Overall this
increases the CNOT count by at most $n$ so it is negligible in the
derivation of the upper bound above.

\subsection{Optimizing the Choice of \texorpdfstring{$L$}{L} and \texorpdfstring{$U$}{U}}

An LU decomposition provides a way to decompose an operator into two
simpler sub-operators. Ideally, the path from the identity to the
target operator given by the concatenation of the paths of two
sub-operators would be close to the optimal path. In our case, it is
very unlikely that an LU decomposition would give such an ideal
decomposition because the structure of triangular operators is too
specific. Nonetheless, the matrices L and U in the LU decomposition
are not unique and we can optimize the factorization by choosing
appropriate $L$ and $U$ operators such that we minimize the overall
complexity of the final reversible circuit. In this section, we report
several strategies to choose the matrices L and U.

If we allow row and column permutations, we can compute the LU
decomposition of an operator $A$ following this
algorithm:
\begin{enumerate} 
\item Choose a row vector $v = A[i,:]$ and a column $j$ for which
  $A[i,j] = 1$.
\item Set the first line of $U$ to $v$ and set the first column of $L$
  to $A[:,j]$.
\item Update the matrix $A$, first by left-applying the gate
  $\text{SWAP}(i,1)$ and right-applying the gate
  $\text{SWAP}(j,1)$. Then add the vector $v$ to every row $A[k,:]$
  where $A[k,1] = 1$. Finally we are left with a matrix $A$ with
  zeroed first column and first row.
\item Repeat the algorithm on this updated $A$ until we obtain the
  null matrix.
\end{enumerate}
At the end of the algorithm we have the relation
\[
  P_1 A P_2 = LU
\]
where $P_1, P_2$ are permutation matrices. $P_1$ is the concatenation
of the SWAP gates that were left-applied to $A$ and $P_2$ is the
concatenation of the SWAP gates that were right-applied to $A$. After
the synthesis of $L$ and $U$ we can commute the gates with $P_2$ and
have
\[
  A = P_1^{-1} P_2^{-1} L'U'
\] 
and we transfer the permutation matrix $P_1^{-1}P_2^{-1}$ to the end
of the total circuit. $L'$ and $U'$ are triangular operators up to a
permutation of their rows and columns.

At each step of the algorithm, we have several choices for the row and
column vectors.  We aim to construct $L$ and $U$ such that their
synthesis will give short circuits. To do so we tried different
strategies :
\begin{itemize}
\item First, if we believe that sparse $L$ and $U$ matrices can be
  implemented with shorter circuits, a first approach consists in
  choosing the row and column with smallest Hamming weights: this is
  the ``sparse'' approach.
\item Secondly, we can adopt a "minimizing cost" approach. If the
  updated matrix $A'$ minimizes a cost function then the synthesis of
  $A'$ will require a shorter circuit. We choose the cost function to
  be the number of ones and we choose among all the possible rows and
  columns the pair that minimizes the number of ones in $A'$.
\end{itemize}
These decompositions transform a general operator $A$ into two
operators $L, U$ with specific structures. Triangular operators are
easier to synthesize but this also limits the optimality of the
solution. Because we force the algorithm to synthesize two matrices
with a specific structure, it is very unlikely that we can perform an
optimized synthesis by going through this forced path. This also gives
insights into the question about the global optimality of our
algorithm: even if we use an optimal algorithm for synthesizing a
triangular operator it is very unlikely that we can get an optimal
solution for the whole synthesis. Our algorithm gives asymptotic
optimal results in the worst case but will be suboptimal for specific
operators. Notably, if the operators can indeed be synthesized with a
small circuit, direct greedy methods have shown encouraging
results. In the next section, we will review the greedy method used in
the literature and propose an extension of this framework to improve
the scalability of this kind of algorithms.

\section{Pathfinding Based Algorithms} \label{cnot_size::pathfinding}

The problem of linear reversible circuit synthesis can fit into a more
generic problem called "pathfinding". In a pathfinding problem, we are
generally given a starting state, a target state, and a set of
operators that dictates the reachable states from a specific
state. Then the goal is to find a path from the starting state to the
goal state and the shorter the path is, the better it is. The
reformulation of linear reversible circuit synthesis into a
pathfinding problem is straightforward: the starting state is the
operator $A$ we want to implement, the goal state is the identity
operator and the available operators are all the elementary row and
column operations.

This field of research in AI has been very active for the last decades
and we can rely on many algorithms to solve our problem. The most
famous one is probably the A* algorithm and all its derivatives (IDA*
\cite{korf1985depth}, Weighted A* \cite{korf1996artificial}, Anytime
A* \cite{likhachev2004ara}, etc.). These algorithms have been
successfully used to solve toy problems (Rubik's cube, Tile puzzles,
Hanoi towers, Top Spin puzzle) but also concrete problems in video
games \cite{DBLP:journals/ijcgt/AlgfoorSK15}, robotics
\cite{likhachev2004ara,DBLP:journals/ijcgt/AlgfoorSK15}, genomics
\cite{DBLP:conf/aaai/ZhouH02,DBLP:journals/jair/HansenZ07,%
DBLP:conf/aaai/YoshizumiMI00,DBLP:journals/tcs/IkedaI99},
planning
\cite{DBLP:journals/ai/BonetG01,DBLP:journals/jair/HansenZ07},
etc. Yet we are quickly faced with some crucial limitations :
\begin{itemize}
\item First the size of the search space grows much faster in our case
  than in any of the cases mentioned above. To give an order of
  magnitude, in \cite{wilt2010comparison} they compare greedy search
  algorithms - a subclass of pathfinding algorithms known for
  providing suboptimal solutions very quickly - and they highlight
  time and memory limitations of some algorithms especially on the
  largest problem they ran on: the 48-puzzle. The 48-puzzle and its
  variations (8-puzzle, 15-puzzle) are sliding puzzles. In the case of
  the 48-puzzle, 48 tiles are arranged in a $7 \times 7$ square with
  one tile missing. The goal of the puzzle is to order the tiles by
  moving the empty space around. For this specific problem the authors
  of \cite{wilt2010comparison} show that the computational time for
  most of the pathfinding algorithms they consider increases
  substantially while providing solutions of poor quality. The number
  of different positions in the 48-puzzle is equal to
  $49!/2 \approx 3 \times 10^{62}$. This is less than the number of
  linear reversible functions on $15$ qubits. As we expect to be able
  to synthesize circuits on $50$ qubits and even more, it is clear
  that we cannot use all the pathfinding algorithms in the literature.
\item Secondly we need a function that will evaluate the distance
  between the current state and the target state. The properties of
  this function, usually called the heuristic, will deeply influence
  the performance of the pathfinding algorithms. Its goal is to guide
  the search by giving priority to more promising states and prune a
  whole part of the search space. More generally there are two sources
  of improvement for a pathfinding algorithm: the accuracy of the
  heuristic function and the management of the path discovery, i.e.,
  what strategy do you employ to explore the search space. If the
  latter can be quite independent of the problem, the former is very
  problem-specific and unfortunately to our knowledge no good
  heuristics exist that satisfy good theoretical properties. For
  instance, for A* to provide an optimal solution we need the
  heuristic to be admissible and consistent, i.e., it must always
  underestimate the length of the optimal path and the value of the
  heuristic should always decrease of at most one after the
  application of one elementary operation. By taking a function that
  always returns 0 we have an admissible and consistent heuristic but
  the search will result in a brute-force search. Generally to obtain
  an admissible heuristic we can consider a simpler version of our
  problem and the optimal solution of this problem gives an
  underestimation of the distance to the target state. In the case of
  linear reversible circuits, the closest simplified subproblem we
  found is the Shortest Linear Straight-Line Program
  \cite{DBLP:journals/joc/BoyarMP13} but this problem is also NP-hard
  and cannot be approximated exactly by a polynomial algorithm.
\end{itemize}

For all these reasons, we believe that using the standard A* algorithm
or its derivatives is a dead end. Instead of that, we focused our work
on greedy best-first search algorithms that can be seen as a cost
minimization/discrete gradient descent approach. It has been shown
that the heuristics that are efficient for this kind of search are
different than the ones for A* algorithms \cite{wilt2016effective}, so
we expect that we can exploit the properties of the matrix $A$ of our
operator to guide our search.

To our knowledge, the cost minimization approach for linear reversible
circuits synthesis has only been investigated in
\cite{schaeffer2014cost}. Yet, with their methods, the authors can
only synthesize circuits on at most $25$ qubits. More crucially, they
did not investigate the behavior of their methods when the resulting
circuits are supposed to be small. According to us, this is a behavior
that has to be considered because as we will see the results of the
methods drastically change depending on the size of the optimal
circuits.

We consider two different heuristic functions to guide our search. The
first one, already used in \cite{schaeffer2014cost}, is the function
\[ h_{\text sum}(A) = \sum_{i,j} a_{i,j}\] that is to say the number
of ones in the matrix $A$. We also propose a new cost function to
solve this problem, it is defined by
\[ h_{\text prod}(A) = \sum_{i=1}^{n} \log\left(\sum_{j=1}^n a_{i,j}
  \right)\] which corresponds to the log of the product of the number
of ones on each row. These two cost functions reach their minimum when
$A$ is a permutation matrix, motivating their use in a cost
minimization process. If the cost function $h_{\text sum}$ seems the
first natural choice, we propose the cost function $h_{\text prod}$
because it gives priority to "almost done" rows. Namely, if one row
has only a few nonzero entries, the minimization process with
$h_{\text prod}$ will treat this row in priority and then it will not
modify it anymore. This enables to avoid a problem which one meets
with the cost function $h_{\text sum}$ where one ends up with a very
sparse matrix but where the rows and columns have few nonzero common
entries. This type of matrix represents a local minimum from which it
can be difficult to escape. With this new cost function, as we put an
additional priority on the rows with few remaining nonzero entries, we
avoid this pitfall.

An important improvement given in \cite{schaeffer2014cost} is to add
the value of the cost function applied to the inverse as well, i.e.,
\[ H_{\text sum}(A) = h_{\text sum}(A) + h_{\text sum}(A^{-1}),\]
\[ H_{\text prod}(A) = h_{\text prod}(A) + h_{\text prod}(A^{-1}).\]
Again, these two new cost functions are minimum when $A$ is a
permutation. As we will see adding the cost value of the inverse
tightens the possible paths and can improve the results because we
escape more easily from local minima during the search.

We now give a few details on our implementation. We directly work on a
cost matrix that stores the impact of each possible CNOT on the cost
function. Namely let $M$ be such matrix, then
$M[i,j] = h(E_{ij}A) - h(A)$ where
$h \in \{ h_{\text sum}, h_{\text prod} \}$. We can define similar
cost matrices for $A^{-1}$ and for column operations. Overall we have
four matrices
$M_{A, \text{row}}, M_{A, \text{col}}, M_{A^{-1}, \text{row}},
M_{A^{-1}, \text{col}}$. Note that $(E_{ij}A)^{-1} = A^{-1}E_{ij}$ so
a row operation $\text{Row}(i,j)$ on $A$ is a column operation
$\text{Col}(j,i)$ on $A^{-1}$. The cost minimization algorithm is
simple if we work with such matrices: at each iteration we look for
the pair $(i,j)$ that minimizes
$\min(M_{A, \text{row}}[i,j] + M_{A^{-1}, \text{col}}[j,i], M_{A,
  \text{col}}[i,j] + M_{A^{-1}, \text{row}}[j,i])$, we update $A$ and
$A^{-1}$ and we pursue the algorithm until $A$ is a permutation
matrix. The advantage of working on cost matrices rather than directly
on $A$ is that the update of the $M$'s is cheaper than updating $A$
and computing the costs. For instance, suppose we apply
$\text{Row}(i,j)$ to $A$, then if $h = h_{\text{sum}}$ the impact on
the cost function of any row operation that does not involve row $j$
will be the same than before the application of
$\text{Row}(i,j)$. This means that we only need to update
$M_{A, \text{Row}}[i,j], M_{A, \text{Row}}[j,i], M_{A^{-1},
  \text{Col}}[i,j], M_{A^{-1}, \text{Col}}[j,i]$ for $i=1..n$. The
updates of $M_{A, \text{Col}}$ and $M_{A^{-1}, \text{Row}}$ are also
simpler because only one element of each column of $A$ has been
modified. However, for our new cost function the updates are not that
simple and it will have an impact on the computational time.

With cost minimization techniques it is not rare to fall into local
minima. In that case we simply select among the best operations a
random one and pursue the search. If the number of iterations exceeds
a certain number then we consider that the algorithm is stuck in a
local minimum and we stop it.

\section{Benchmarks} \label{cnot_size::bench}

This section presents our experimental results. We have the following
algorithms to benchmark:
\begin{itemize}
\item GreedyGE from Sections~\ref{cnot_size::greedyge} and
  \ref{cnot_size::extension},
\item cost minimization techniques from
  Section~\ref{cnot_size::pathfinding},
\end{itemize}

The state-of-the-art algorithms are the following: 
\begin{itemize}
\item The PMH algorithm \algoone,
\item The algorithm based on the syndrome decoding problem
  from~\cite{DBLP:conf/rc/BrugiereBVMA20}.
\end{itemize}

Two kinds of datasets are used to benchmark our algorithms:
\begin{itemize}
\item First, a set of random operators. The test on random operators
  gives an overview of the average performance of our algorithm. We
  generate random operators by creating random CNOT circuits. Our
  routine takes two inputs: the number of qubit $n$ and the number of
  CNOT gates $k$ in the random circuit. Each CNOT is randomly placed
  by selecting a random control and a random target and the simulation
  of the circuit gives a random operator. Empirically we noticed that
  when $k$ is sufficiently large --- $k=n^2$ is enough --- then the
  operators generated have strong probability to represent the worst
  case scenarii.
\item Secondly, a set of reversible functions, given as circuits,
  taken from Matthew Amy's github repository \cite{meamy}. This
  experiment shows how our algorithms can optimize useful quantum
  algorithms in the literature like the Galois Field multipliers,
  integer addition, Hamming coding functions, the hidden weighted bit
  functions, etc.
\end{itemize}

To evaluate the performance of our algorithms for the random set, two
types of experiments are conducted:
\begin{enumerate}
\item a worst-case asymptotic experiment, namely for increasing
  problem sizes $n$ we generate circuits with $n^2$ gates and we
  compute the average number of gates for each problem size. This
  experiment reveals the asymptotic behavior of the algorithms and
  gives insights about strict upper bounds on their performance.
\item a close-to-optimal experiment, namely for one specific problem
  size we generate operators with different number of gates to show
  how close to optimal our algorithms are if the optimal circuits are
  expected to be smaller than the worst case.
\end{enumerate}

All our algorithms are implemented in Julia \cite{bezanson2017julia}
and executed on the ATOS QLM (Quantum Learning Machine) whose
processor is an Intel Xeon(R) E7-8890 v4 at 2.4 GHz.

\subsection{Random Operators} \label{bench::size_unconstrained}

\subsubsection{GreedyGE}

First, we present the worst-case asymptotic experiment with the
following algorithms:
\begin{itemize}
\item GreedyGE with standard LU decomposition,
\item Syndrome decoding with the cost minimization heuristic with
  unlimited width and depth 1,
\item Syndrome decoding with the integer programming solver (Coin-or
  branch and cut solver),
\item Syndrome decoding with the cost minimization heuristic
  (width=Inf, depth=1) and $50$ random changes of basis, the
  ``Information Set Decoding'' (ISD) case.
\item Syndrome decoding with the cost minimization heuristic with
  width 60 and depth 2,
\item Syndrome decoding with the cost minimization heuristic with
  width 15 and depth 3,
\end{itemize}

\begin{figure}
\center
\includegraphics[scale=0.5]{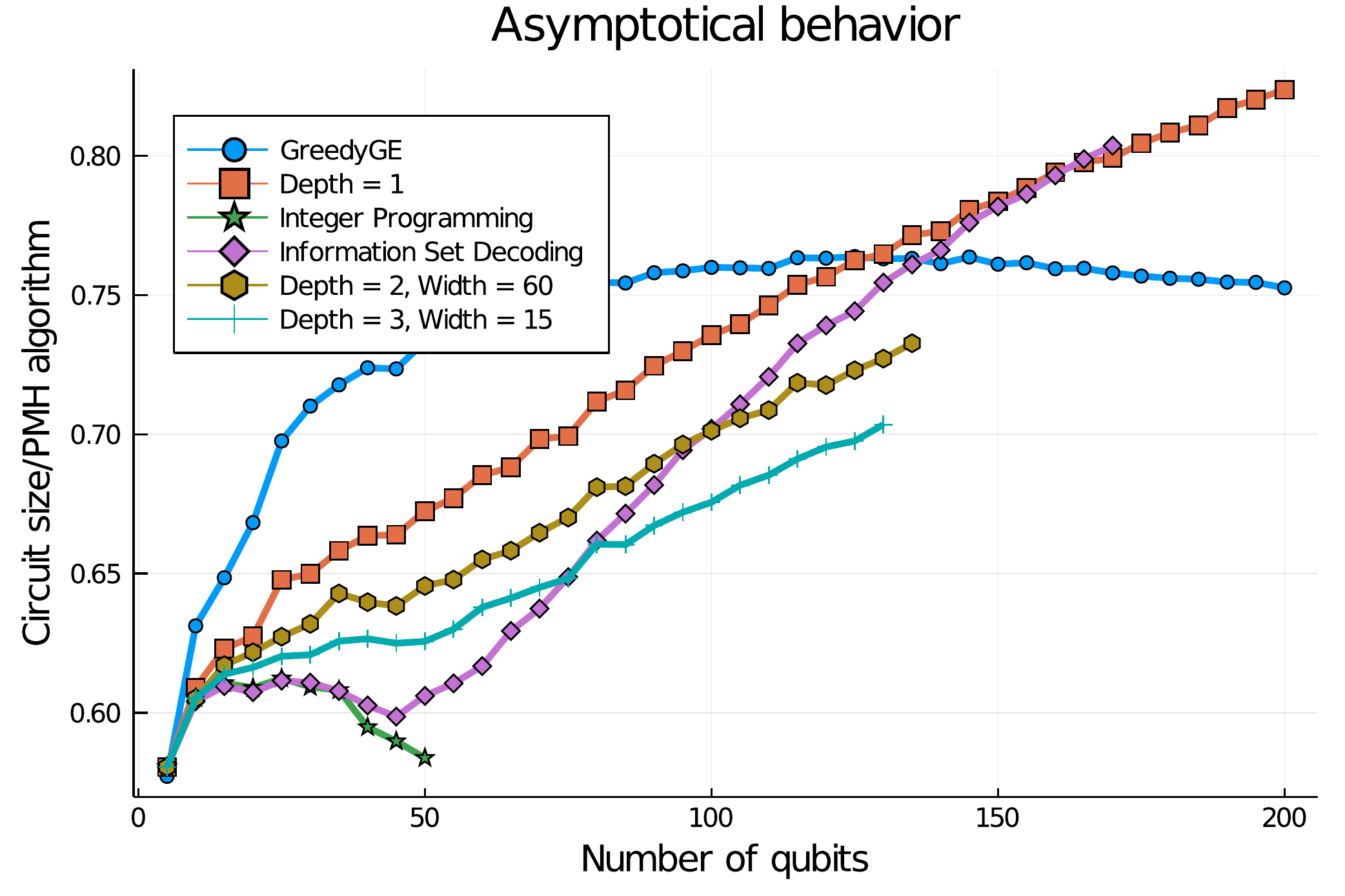}
\def\mycap{Average performance of GreedyGE and the Syndrome Decoding
  based algorithms versus the PMH algorithm.}%
\caption{\mycap}
\label{bench::size_average}
\end{figure}

The results are given in Fig~\ref{bench::size_average}. For the sake
of clarity, instead of showing the circuit size we plot the ratio
between the circuit size given by our method and the circuit size
given by the PMH algorithm. So if the ratio is below $1$ this means
that we outperform the state-of-the art method \algoone. We also have
a better view of which algorithm is better in which qubit range. We
stopped the calculations when the running time was too large for
producing benchmarks within several hours.

Overall, for the considered range of qubits, GreedyGE outperforms
{\algoone}.  As the number of qubits increases the syndrome decoding
based algorithm performs worse and GreedyGE eventually produces the
shortest circuits when $n > 120$. Notably, the average gain over the
PMH algorithm stabilizes around 25\%. On the other hand the syndrome
decoding based algorithms performances deteriorate as $n$ increases or
the algorithm cannot terminate when the heuristic is too costly.
Therefore, for large $n$, GreedyGE is the best method so far.
GreedyGE is also fast: we compare the computational time of GreedyGE
against the standard Gaussian elimination and the PMH algorithm. The
results are given in Table~\ref{bench::time}.  Not only GreedyGE
produces the smallest circuits but in addition it is the fastest
method to our knowledge.

\begin{table}
  \caption{Computational time of GreedyGE vs PMH and standard Gaussian
    elimination algorithms.}
\center
\begin{tabular}{c|ccccc}
\toprule
\multirow{2}{*}{Algorithm} & \multicolumn{5}{c}{Average computational time (s)} \\
                           & $n=10$ & $n=100$ & $n=500$ & $n=1000$ & $n=5000$ \\
\midrule                           
Gaussian elimination       & 0.00004     & 0.00125      &  0.14     &  1.2      &  140  \\
PMH algorithm              & 0.00015     & 0.006        &  0.75     &  7        &  980  \\
GreedyGE                   & 0.00015     & 0.0035       &  0.1      &  0.6      &  45  \\
\bottomrule
\end{tabular}
\label{bench::time}
\end{table}

Next, we show the impact of the choice of the LU decomposition. We
provide results with GreedyGE considering it can be directly
transposed to the syndrome decoding based algorithms as well. We
perform the close-to-optimal experiment with the GreedyGE and three
different LU decompositions:
\begin{itemize}
\item a standard LU decomposition for matrix in $\mathbb{F}_2$, taken
  from the Julia package Nemo \cite{nemo},
\item the LU decomposition algorithm with the "sparse" strategy,
\item the LU decomposition algorithm with the "cost minimization"
  strategy.
\end{itemize}

\begin{figure}
  \center \includegraphics[scale=0.5]{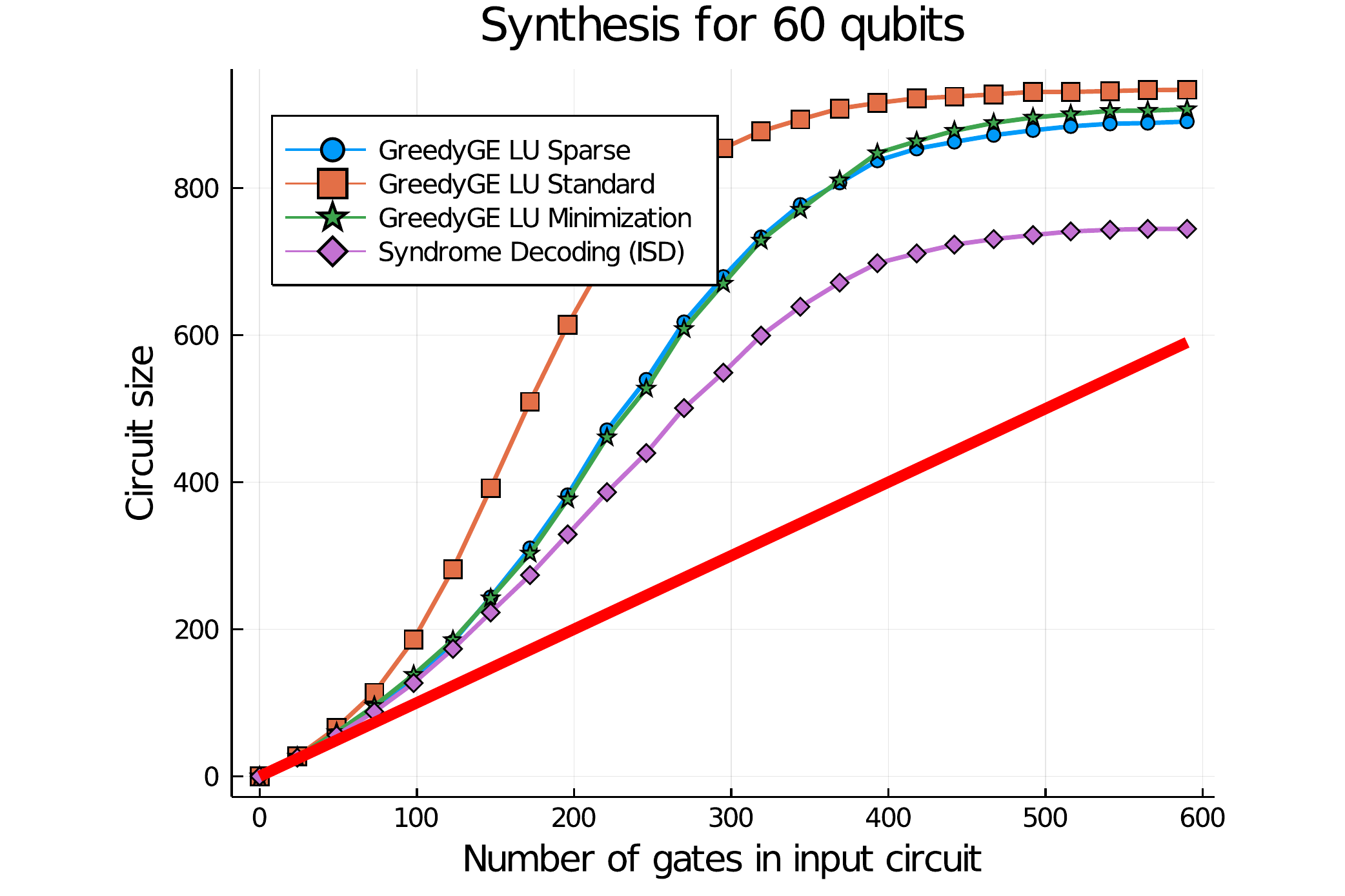}
  \def\mycapt{Performance of GreedyGE with different LU decompositions
    vs Syndrome Decoding on 60 qubits for different input circuit
    sizes.}%
\caption{\mycapt}
\label{bench::size_60qubits}
\end{figure}

The experiment is done on $60$ qubits. We also add the best method for
this problem size: the syndrome decoding based algorithm with the
"Information Set Decoding" strategy.  The results are given in
Fig~\ref{bench::size_60qubits}. Computing more efficient LU
decompositions has almost no effect on the worst-case results but
provides some improvements when the input circuits are smaller. There
is no significant difference between the two strategies "sparse" and
"cost minimization" in terms of circuit sizes but the running time is
much lower for the sparse approach so we would privilege it. This
improvement also benefits the syndrome decoding based method as it
also relies on an LU decomposition. On the other hand we also see that
there is no obvious advantage of using GreedyGE instead of the
syndrome decoding based methods when the optimal circuit is expected
to be small.

\subsubsection{Path Finding Methods}

We now evaluate the performance of purely greedy methods. We remind
that we have four cost functions to study:
\begin{itemize}
  \item $h_{\text sum}(A) = \sum_{i,j} a_{i,j}$,
  \item $h_{\text prod}(A) = \sum_{i=1}^{n} \log\left(\sum_{j=1}^n a_{i,j} \right)$,
  \item $H_{\text sum}(A) = h_{\text sum}(A) + h_{\text sum}(A^{-1})$,
  \item $H_{\text prod}(A) = h_{\text prod}(A) + h_{\text prod}(A^{-1})$.
\end{itemize}

\begin{figure}
\centering
\begin{subfigure}{\textwidth}
  \centering
  \includegraphics[scale=0.5]{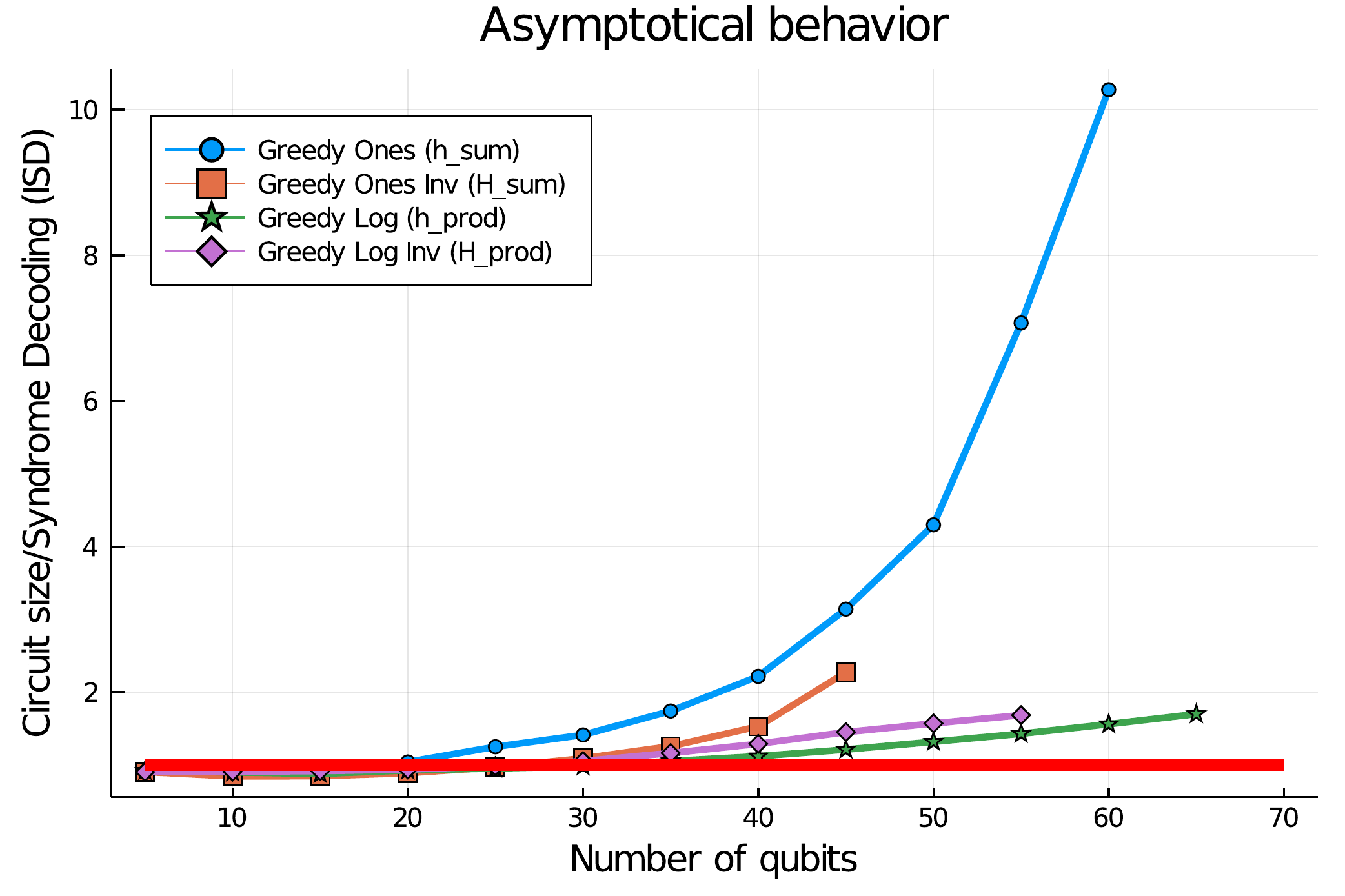}
  \caption{Average performance of \{$h_{\text sum},h_{\text prod},H_{\text sum},H_{\text prod}$\} vs Syndrome decoding (Information Set Decoding).}
  \label{bench::asymptotic_bench_greedy}
\end{subfigure}
\\
\begin{subfigure}{\textwidth}
  \centering
  \includegraphics[scale=0.5]{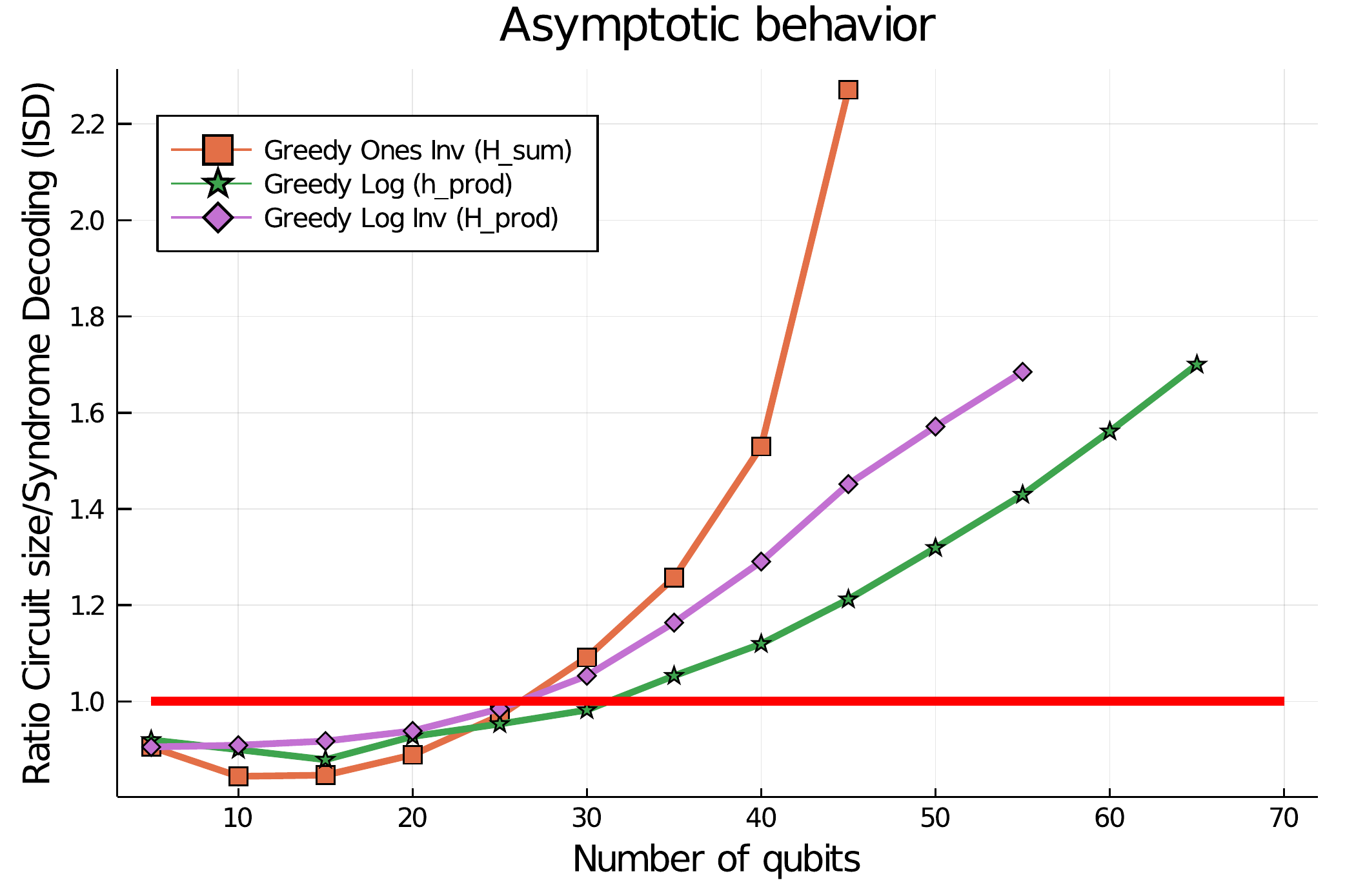}
\caption{Same graph as in Fig~\ref{bench::asymptotic_bench_greedy} without $h_{\text sum}$.}
\label{bench::asymptotic_bench_greedy2}
\end{subfigure}
\caption{Average performance of Cost minimization techniques vs Syndrome decoding (Information Set Decoding).}
\label{bench::size_average_greedy}
\end{figure}

First, we show the asymptotic behavior of those four methods. The
results are given in Fig~\ref{bench::size_average_greedy}.  The first
graph in Fig~\ref{bench::asymptotic_bench_greedy} shows that the cost
function $h_{\text sum}$ does not scale well with the number of qubits
so we decide to remove it for clarity. The new graph is given in
Fig~\ref{bench::asymptotic_bench_greedy2}. We notice two things:
first, the cost function $H_{\text prod}$ always underperforms
$h_{\text prod}$, secondly both $H_{\text sum}$ and $h_{\text prod}$
outperform our syndrome decoding based algorithm for small problem
sizes ($n < 30$) but with an advantage for $H_{\text sum}$ when
$n < 25$. There is a thin window --- between $25$ and $30$ qubits ---
where it is preferable to use $h_{\text prod}$ instead of
$H_{\text sum}$.

\begin{figure}
  \center
  \includegraphics[scale=0.5]{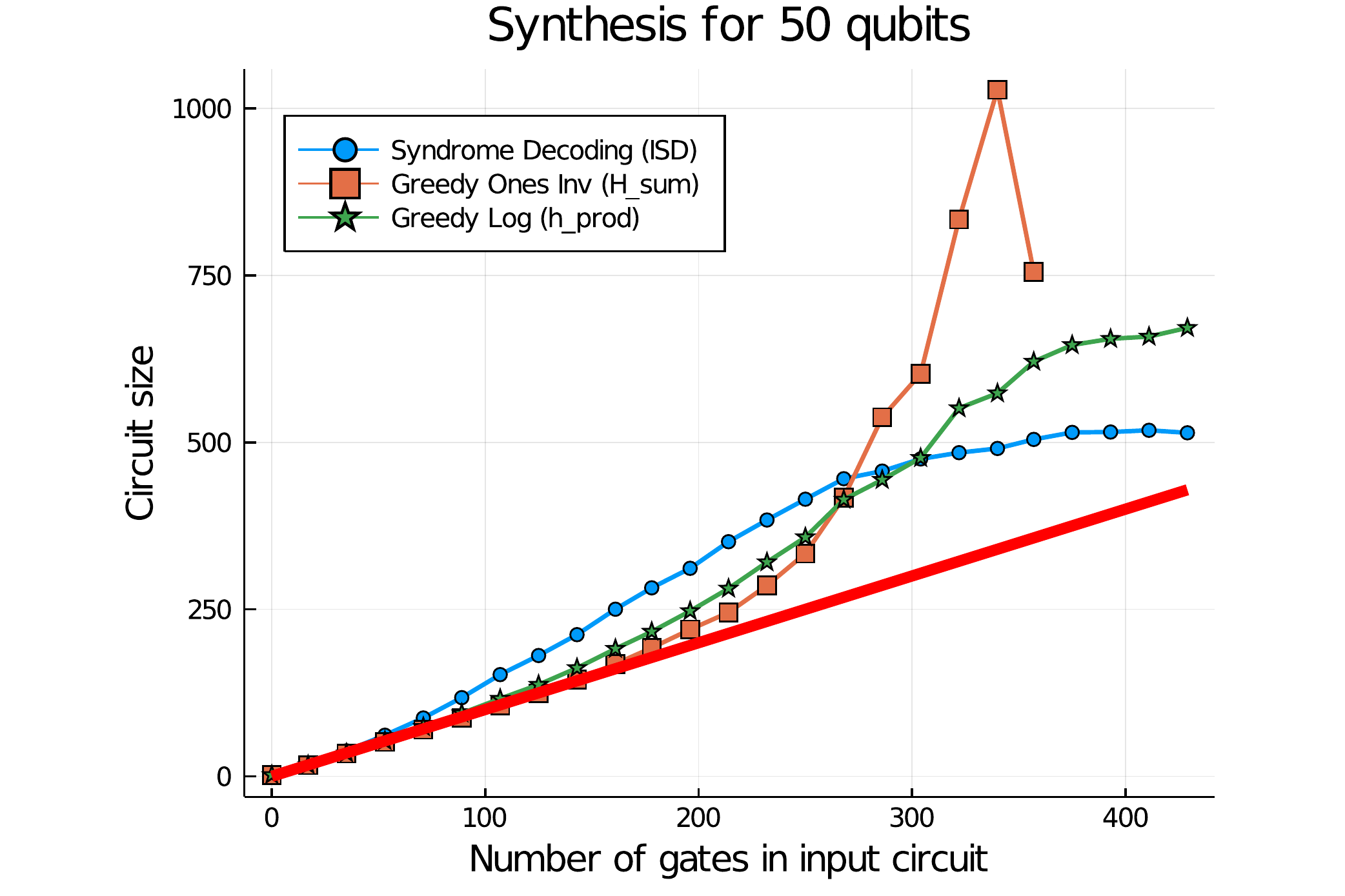}
\caption{Performance of Cost minimization techniques vs Syndrome Decoding on 50 qubits for different input circuit sizes.}
\label{bench::size_50qubits}
\end{figure}

The results given in Fig~\ref{bench::size_average_greedy} are mainly
there to discredit two of the four cost functions: $h_{\text sum}$ and
$H_{\text prod}$.  The result that really interests us is that of the
close-to-optimal experiment, here on $50$ qubits and given in
Fig~\ref{bench::size_50qubits}.  For this problem size on the worst
case the syndrome decoding based algorithm outperforms our greedy
methods. But when we are not in the worst case the results are
completely different: the greedy methods follow more faithfully the
bound $y=x$ than the syndrome decoding based algorithm. We have to
wait input circuits of size $300$ for the syndrome decoding algorithm
to be better. The greedy algorithm based on the cost function
$h_{\text prod}$ produces much more stable results than the one with
the cost function $H_{\text sum}$. For the cost function
$H_{\text sum}$ the variance is extremely large because the method
struggles finding a global minimum.  However, it is $H_{\text sum}$
that produces the best results most of the time in the range of size
$0-300$. Again there is a thin window where it may be relevant to use
$h_{\text prod}$ instead of $H_{\text sum}$.

\subsubsection{Conclusion: Combining the Methods}

Each method has its own range of validity:
\begin{itemize}
\item GreedyGE is suited when the number of qubits exceeds $120$, it
  outperforms the PMH algorithm and our other algorithms in both
  circuit size and computational time.
\item The syndrome decoding based algorithms must be used for
  intermediate problem sizes ($30 < n < 120$) with the best possible
  solver of the syndrome decoding problem (in order: integer
  programming, Information Set Decoding if $n < 70$, maximum depth
  otherwise). Both GreedyGE and the family of syndrome decoding based
  algorithms are state of the art in a worst case scenario.
\item When the output circuit is expected to be small, or when
  $n < 30$, then direct greedy methods have shown to produce the best
  results. The proposal of a new cost function paid off as this offers
  a more scalable direct greedy search. However, the moment when our
  new custom function outperforms the one proposed in
  \cite{schaeffer2014cost} coincides with the moment when the syndrome
  decoding based algorithm also outperforms the direct greedy
  search. More investigation needs to be done to clarify which cost
  function should be preferred and when.
\end{itemize}

Given the variety of methods, each providing the best results in
particular cases, the best method when trying to synthesize an
operator would be simply to test each method and to keep the best
result. Except for the direct greedy methods, the computational time
of each method is well understood. Given a specific problem size, it
is easy to know which method can be used and how long it will take. It
is more delicate with the direct greedy methods, we have to set a
limit in the maximum number of iterations before considering that the
search will not converge to a solution.

\subsection{Benchmarks on Reversible Functions} \label{cnot_size::bench_reversible}

We apply our CNOT circuit synthesis to the optimization of quantum
algorithms. A library to evaluate the quality of a quantum compiler is
a library of reversible functions consisting of adders, Hamming coding
functions, multipliers, etc. This library is a standard benchmark for
circuit optimization algorithms and is used in recent articles about
quantum circuit optimization \cite{2058-9565-4-1-015004,
  DBLP:journals/tcad/AmyMM14,amy2018controlled,Continuous}. However,
we must be aware that our results can only be compared to those using
the same methodology. There are many other methods for the synthesis
and optimization of reversible circuits \cite{saeedi2013synthesis},
notably for the synthesis of oracles
\cite{DBLP:conf/iccad/MeuliSCRM19, DBLP:journals/corr/abs-2005-00211},
which we do not take into account here. The aim of this section is to
show that our methods can bring promising results on specific circuits
and that it will be interesting to apply it on other classes of
circuits, but it will also be necessary to put it in perspective with
other methods.

The original circuits are from Matthew Amy's github Feynman repository
\cite{meamy}.

Recent quantum compilers mainly focused on the T-count or T-depth
optimization of quantum circuits but these optimizations often lead to
an increased CNOT count \cite{2058-9565-4-1-015004,
  DBLP:journals/tcad/AmyMM14}.  Although it is experimentally more
costly to implement a T gate, the total number of CNOT gates in a
quantum circuits should not be too large otherwise the CNOT cost of
the total circuit will not be negligible. In practice it may even be
possible that the CNOT cost represents the most costly part of a
quantum circuit \cite{DBLP:journals/qic/Maslov16} and it cannot be
neglected.

In \cite{amy2018controlled} the algorithm Gray-Synth for the
optimization of the CNOT count in CNOT+T circuits is proposed. It
efficiently reduces the total number of CNOT gates but it comes at the
price of an increased T-depth. Here we show that, using our methods
for CNOT circuit synthesis, one can reduce significantly the CNOT
count without paying the price of increasing the T depth. We used the
C++ implementation of Matthew Amy's Tpar algorithm to produce
optimized circuits with low T-count and T-depth. We post process the
circuits by re-synthesizing any chunk of purely CNOT circuits. It is
possible to further optimize the synthesis of the CNOT+T circuits,
using for instance the TODD optimizer \cite{2058-9565-4-1-015004} but
as we are concerned about showing the impact of our method on the CNOT
count and the T-depth we did not pursue the optimization to have a
lower T-count so that we can directly compare against the results from
\cite{amy2018controlled}.

The results are given in Table~\ref{table_results}. For each
reversible function we provide the results (T-count, T-depth, Total
depth) of different methods implementing it. Namely, the original
circuits, the circuits optimized solely with the Tpar algorithm and
GraySynth as the state-of-the-art methods and the circuits optimized
with the Tpar method and our own method, i.e., a mix of GreedyGE, the
syndrome decoding based methods and the purely greedy methods. For
each CNOT circuit we keep the best result encountered. To generate the
Gray-Synth results we used the Haskell implementation, still from
\cite{meamy}. The savings given always compare against the circuits
given by solely the Tpar algorithm. We notice several interesting
points:
\begin{itemize}
\item we cannot decrease the CNOT count as much as with GraySynth, but
  we still manage to decrease the CNOT count consistently and above
  all we do not modify the T-depth. Our circuits therefore represent a
  tradeoff between the T-cost the CNOT-cost.
\item We also considerably reduce the total depth of the circuits,
  making their execution faster on NISQ architectures. This is
  explained by the fact that most of the CNOT circuits involved in
  those reversible functions are elementary operators, requiring only
  a few CNOTs to be implemented. Therefore the size and the depth are
  very close and greedy methods find close-to-optimal implementations.
\end{itemize}

To illustrate our last statement, we kept track of the best algorithm
that was used to compute each CNOT sub-circuit that appeared in the
synthesis of one reversible function.  The results are given in
Table~\ref{bench::frequency}. Most of the time, both purely greedy
methods and the syndrome decoding based methods provide the best
results. According to us, this highlights the simplicity of the
operators to synthesize otherwise there would be many more cases where
we would observe a difference of even one CNOT. Overall, for this
library of reversible functions, the use of methods other than the
pure greedy methods is useless. This is not surprising as the number
of qubits never exceeds $40$ except for three operators. For such
small problem sizes, we saw that pure greedy methods always outperform
the other methods, in addition to the fact that we have strong
suspicions that the majority of the operators encountered are easy to
synthesize. We note the interest of using the cost function
$h_{\text{prod}}$ as an alternative because in some cases it allows to
synthesize operators more efficiently.

\begin{sidewaystable*}
\centering
\caption{CNOT optimization of a library of reversible functions with
  several CNOT circuits synthesis methods.}
\scalebox{.62}{
\begin{tabular}{lccccccccccccccccc} \toprule
      \multirow{2}{*}{Function} & \multirow{2}{*}{\#$n$} & \multicolumn{4}{c}{Original} & \multicolumn{4}{c}{Tpar} & \multicolumn{4}{c}{Tpar + GraySynth} & \multicolumn{4}{c}{Tpar + CNOT size opt.} \\
      \cmidrule(lr){3-6} \cmidrule(lr){7-10} \cmidrule(lr){11-14} \cmidrule{15-18}
               &     &  T-count & T-depth & CNOT count & Depth &  T-count & T-depth & CNOT count & Depth &  T-count & T-depth & CNOT count & Depth &  T-count & T-depth & CNOT count & Depth \\
      \ \\
Adder\_8& $ 24$ & $399$ & $69$ & $466$ & $223$ & $213$ & $30$ & $741$ & $302$ & $215$ & $74$ ( $147\% $)& $399$ ($ -46\% $) & $257$ ( $ -15\% $) & $213$ & $30$ & $491$ ( $ -34\% $) & $204$ ( -32\%) \\
barenco\_tof\_10& $ 19$ & $224$ & $96$ & $224$ & $288$ & $100$ & $43$ & $332$ & $272$ & $100$ & $81$ ( $88\% $)& $146$ ($ -56\% $) & $264$ ( $ -3\% $) & $100$ & $43$ & $188$ ( $ -43\% $) & $217$ ( -20\%) \\
barenco\_tof\_3& $ 5$ & $28$ & $12$ & $28$ & $36$ & $16$ & $8$ & $52$ & $60$ & $16$ & $16$ ( $100\% $)& $20$ ($ -62\% $) & $40$ ( $ -33\% $) & $16$ & $8$ & $26$ ( $ -50\% $) & $34$ ( -43\%) \\
barenco\_tof\_4& $ 7$ & $56$ & $24$ & $56$ & $72$ & $28$ & $13$ & $96$ & $96$ & $28$ & $26$ ( $100\% $)& $38$ ($ -60\% $) & $64$ ( $ -33\% $) & $28$ & $13$ & $50$ ( $ -48\% $) & $61$ ( -36\%)  \\
barenco\_tof\_5& $ 9$ & $84$ & $36$ & $84$ & $108$ & $40$ & $18$ & $134$ & $123$ & $40$ & $34$ ( $89\% $)& $56$ ($ -58\% $) & $88$ ( $ -28\% $) & $40$ & $18$ & $73$ ( $ -46\% $) & $89$ ( -28\%) \\
csla\_mux\_3& $ 15$ & $70$ & $21$ & $90$ & $67$ & $62$ & $8$ & $379$ & $210$ & $62$ & $28$ ( $250\% $)& $115$ ($ -70\% $) & $91$ ( $ -57\% $) & $62$ & $8$ & $187$ ( $ -51\% $) & $81$ ( -61\%) \\
csum\_mux\_9& $ 30$ & $196$ & $18$ & $196$ & $59$ & $84$ & $6$ & $366$ & $153$ & $84$ & $16$ ( $167\% $)& $160$ ($ -56\% $) & $78$ ( $ -49\% $) & $84$ & $6$ & $179$ ( $ -51\% $) & $70$ ( -54\%)  \\
cycle\_17\_3& $ 35$ & $4739$ & $2001$ & $4742$ & $5974$ & $1944$ & $562$ & $6608$ & $5231$ & $1955$ & $1857$ ( $230\% $)& $3040$ ($ -54\% $) & $5698$ ( $ 9\% $) & $1944$ & $562$ & $4267$ ( $ -35\% $) & $4229$ ( -19\%)  \\
GF($2^{10}$)\_mult& $ 30$ & $700$ & $108$ & $709$ & $290$ & $410$ & $16$ & $2206$ & $1026$ & $410$ & $109$ ( $581\% $)& $648$ ($ -71\% $) & $324$ ( $ -68\% $) & $410$ & $16$ & $955$ ( $ -57\% $) & $307$ ( -70\%)  \\
GF($2^{16}$)\_mult& $ 48$ & $1792$ & $180$ & $1837$ & $489$ & $1040$ & $24$ & $6724$ & $2551$ & $1040$ & $585$ ( $2338\% $)& $1691$ ($ -75\% $) & $1488$ ( $ -42\% $) & $1040$ & $24$ & $2542$ ( $ -62\% $) & $678$ ( -73\%) \\
GF($2^{32}$)\_mult& $ 96$ & $7168$ & $372$ & $7292$ & $1001$ & $4128$ & $47$ & $34244$ & $11520$ & $4128$ & $2190$ ( $4560\% $)& $6636$ ($ -81\% $) & $5391$ ( $ -53\% $) & $4128$ & $47$ & $10953$ ( $ -68\% $) & $2113$ ( -82\%)  \\
GF($2^4$)\_mult& $ 12$ & $112$ & $36$ & $115$ & $99$ & $68$ & $6$ & $307$ & $173$ & $68$ & $39$ ( $550\% $)& $106$ ($ -65\% $) & $117$ ( $ -32\% $) & $68$ & $6$ & $135$ ( $ -56\% $) & $80$ ( -54\%)  \\
GF($2^5$)\_mult& $ 15$ & $175$ & $48$ & $179$ & $130$ & $115$ & $9$ & $502$ & $259$ & $115$ & $51$ ( $467\% $)& $166$ ($ -67\% $) & $152$ ( $ -41\% $) & $115$ & $9$ & $209$ ( $ -58\% $) & $107$ ( -59\%) \\
GF($2^6$)\_mult& $ 18$ & $252$ & $60$ & $257$ & $163$ & $150$ & $9$ & $660$ & $350$ & $150$ & $63$ ( $600\% $)& $235$ ($ -64\% $) & $189$ ( $ -46\% $) & $150$ & $9$ & $310$ ( $ -53\% $) & $137$ ( -61\%)  \\
GF($2^7$)\_mult& $ 21$ & $343$ & $72$ & $349$ & $195$ & $217$ & $12$ & $996$ & $490$ & $217$ & $75$ ( $525\% $)& $319$ ($ -68\% $) & $224$ ( $ -54\% $) & $217$ & $12$ & $442$ ( $ -56\% $) & $184$ ( -62\%)  \\
GF($2^8$)\_mult& $ 24$ & $448$ & $84$ & $469$ & $233$ & $264$ & $13$ & $1254$ & $619$ & $264$ & $87$ ( $569\% $)& $428$ ($ -66\% $) & $266$ ( $ -57\% $) & $264$ & $13$ & $588$ ( $ -53\% $) & $224$ ( -64\%)  \\
GF($2^9$)\_mult& $ 27$ & $567$ & $96$ & $575$ & $258$ & $351$ & $15$ & $1712$ & $810$ & $351$ & $95$ ( $533\% $)& $526$ ($ -69\% $) & $283$ ( $ -65\% $) & $351$ & $15$ & $753$ ( $ -56\% $) & $266$ ( -67\%)  \\
grover\_5& $ 9$ & $336$ & $144$ & $336$ & $457$ & $154$ & $51$ & $499$ & $477$ & $154$ & $130$ ( $155\% $)& $232$ ($ -54\% $) & $440$ ( $ -8\% $) & $154$ & $51$ & $331$ ( $ -34\% $) & $384$ ( -19\%)  \\
ham15-high& $ 20$ & $2457$ & $996$ & $2500$ & $3026$ & $1019$ & $380$ & $3427$ & $2956$ & $1019$ & $839$ ( $121\% $)& $1588$ ($ -54\% $) & $2583$ ( $ -13\% $) & $1019$ & $380$ & $2183$ ( $ -36\% $) & $2226$ ( -25\%)  \\
ham15-low& $ 17$ & $161$ & $69$ & $259$ & $263$ & $97$ & $33$ & $471$ & $360$ & $97$ & $83$ ( $152\% $)& $221$ ($ -53\% $) & $271$ ( $ -25\% $) & $97$ & $33$ & $280$ ( $ -41\% $) & $223$ ( -38\%)  \\
ham15-med& $ 17$ & $574$ & $240$ & $616$ & $750$ & $230$ & $84$ & $759$ & $682$ & $230$ & $194$ ( $131\% $)& $368$ ($ -52\% $) & $622$ ( $ -9\% $) & $230$ & $84$ & $481$ ( $ -37\% $) & $506$ ( -26\%)  \\
hwb6& $ 7$ & $105$ & $45$ & $131$ & $152$ & $75$ & $24$ & $270$ & $248$ & $75$ & $63$ ( $162\% $)& $111$ ($ -59\% $) & $188$ ( $ -24\% $) & $75$ & $24$ & $172$ ( $ -36\% $) & $164$ ( -34\%)  \\
hwb8& $ 12$ & $5887$ & $2139$ & $7970$ & $7956$ & $3531$ & $860$ & $22670$ & $15838$ & $3531$ & $2752$ ( $220\% $)& $6841$ ($ -70\% $) & $9338$ ( $ -41\% $) & $3531$ & $860$ & $13353$ ( $ -41\% $) & $9261$ ( -42\%)  \\
mod5\_4& $ 5$ & $28$ & $12$ & $32$ & $41$ & $16$ & $6$ & $48$ & $57$ & $16$ & $15$ ( $150\% $)& $28$ ($ -42\% $) & $51$ ( $ -11\% $) & $16$ & $6$ & $32$ ( $ -33\% $) & $39$ ( -32\%) \\
mod\_adder\_1024& $ 28$ & $1995$ & $831$ & $2005$ & $2503$ & $1011$ & $258$ & $3650$ & $2560$ & $1011$ & $864$ ( $235\% $)& $1390$ ($ -62\% $) & $2474$ ( $ -3\% $) & $1011$ & $258$ & $2369$ ( $ -35\% $) & $1913$ ( -25\%) \\
mod\_adder\_1048576& $ 58$ & $17290$ & $7292$ & $17310$ & $21807$ & $7298$ & $1927$ & $29794$ & $19975$ & $7323$ & $6577$ ( $241\% $)& $11080$ ($ -63\% $) & $20089$ ( $ 1\% $) & $7298$ & $1927$ & $19122$ ( $ -36\% $) & $15286$ ( -23\%)  \\
mod\_mult\_55& $ 9$ & $49$ & $15$ & $55$ & $50$ & $35$ & $7$ & $106$ & $75$ & $35$ & $20$ ( $186\% $)& $40$ ($ -62\% $) & $52$ ( $ -31\% $) & $35$ & $7$ & $73$ ( $ -31\% $) & $61$ ( -19\%)  \\
mod\_red\_21& $ 11$ & $119$ & $48$ & $122$ & $158$ & $73$ & $25$ & $223$ & $207$ & $73$ & $59$ ( $136\% $)& $102$ ($ -54\% $) & $179$ ( $ -14\% $) & $73$ & $25$ & $136$ ( $ -39\% $) & $144$ ( -30\%)  \\
qcla\_adder\_10& $ 36$ & $238$ & $24$ & $267$ & $73$ & $162$ & $13$ & $648$ & $195$ & $162$ & $33$ ( $154\% $)& $225$ ($ -65\% $) & $94$ ( $ -52\% $) & $162$ & $13$ & $362$ ( $ -44\% $) & $86$ ( -56\%)  \\
qcla\_com\_7& $ 24$ & $203$ & $27$ & $215$ & $81$ & $94$ & $12$ & $371$ & $154$ & $94$ & $31$ ( $158\% $)& $150$ ($ -60\% $) & $89$ ( $ -42\% $) & $94$ & $12$ & $202$ ( $ -46\% $) & $76$ ( -51\%)  \\
qcla\_mod\_7& $ 26$ & $413$ & $66$ & $441$ & $197$ & $231$ & $28$ & $813$ & $296$ & $237$ & $67$ ( $139\% $)& $386$ ($ -53\% $) & $212$ ( $ -28\% $) & $231$ & $28$ & $479$ ( $ -41\% $) & $188$ ( -36\%)  \\
qft\_4& $ 5$ & $69$ & $48$ & $48$ & $142$ & $67$ & $44$ & $96$ & $185$ & $67$ & $60$ ( $36\% $)& $48$ ($ -50\% $) & $163$ ( $ -12\% $) & $67$ & $44$ & $56$ ( $ -42\% $) & $150$ ( -19\%)  \\
rc\_adder\_6& $ 14$ & $77$ & $33$ & $104$ & $104$ & $47$ & $22$ & $165$ & $157$ & $47$ & $39$ ( $77\% $)& $79$ ($ -52\% $) & $111$ ( $ -29\% $) & $47$ & $22$ & $100$ ( $ -39\% $) & $103$ ( -34\%)  \\
tof\_10& $ 19$ & $119$ & $51$ & $119$ & $153$ & $71$ & $27$ & $236$ & $190$ & $71$ & $61$ ( $126\% $)& $86$ ($ -64\% $) & $172$ ( $ -9\% $) & $71$ & $27$ & $130$ ( $ -45\% $) & $143$ ( -25\%) \\
tof\_3& $ 5$ & $21$ & $9$ & $21$ & $27$ & $15$ & $6$ & $35$ & $46$ & $15$ & $12$ ( $100\% $)& $16$ ($ -54\% $) & $32$ ( $ -30\% $) & $15$ & $6$ & $21$ ( $ -40\% $) & $31$ ( -33\%)  \\
tof\_4& $ 7$ & $35$ & $15$ & $35$ & $45$ & $23$ & $9$ & $63$ & $71$ & $23$ & $18$ ( $100\% $)& $26$ ($ -59\% $) & $51$ ( $ -28\% $) & $23$ & $9$ & $37$ ( $ -41\% $) & $49$ ( -31\%)  \\
tof\_5& $ 9$ & $49$ & $21$ & $49$ & $63$ & $31$ & $12$ & $97$ & $104$ & $31$ & $24$ ( $100\% $)& $36$ ($ -63\% $) & $70$ ( $ -33\% $) & $31$ & $12$ & $50$ ( $ -48\% $) & $67$ ( -36\%)  \\
vbe\_adder\_3& $ 10$ & $70$ & $24$ & $80$ & $79$ & $24$ & $9$ & $120$ & $88$ & $24$ & $18$ ( $100\% $)& $54$ ($ -55\% $) & $71$ ( $ -19\% $) & $24$ & $8$ & $61$ ( $ -49\% $) & $45$ ( -49\%)  \\
\midrule
      Mean savings (compared to Tpar) &&&&&&&&&&&+391.39\%&-60.21\%&-29.66\%&&&-45.03\%&-41.26\%\\
      Maximum savings (compared to Tpar) &&&&&&&&&&&+36\%&-81\%&-68\%&&&-68\%&-82\%\\
      Minimum savings (compared to Tpar) &&&&&&&&&&&+4560\%&-42\%&+9\%&&&-31\%&-19\%\\
      \bottomrule
      \end{tabular}

}
\label{table_results}
\end{sidewaystable*}

\begin{sidewaystable*}
\centering
 \caption[Frequency of best performance of each algorithm during the optimization of reversible circuits.]{Frequency of best performance of each algorithm during the optimization of reversible circuits. For each algorithm, the first column gives the number of times it has returned the best result (possibly other algorithms returned circuits of same size). The second column reports the number of times it was the only one to provide the best possible circuit.}
\scalebox{0.6}{
\begin{tabular}{lcccccccccccccccc}
\toprule
\multirow{2}{*}{Function} & \multirow{2}{*}{\#$n$} & \multirow{2}{*}{\makecell{\#CNOT \\ sub-circuits}} & \multicolumn{2}{c}{PMH} & \multicolumn{2}{c}{GreedyGE} & \multicolumn{2}{c}{Syndrome (Information Set Decoding)} & \multicolumn{2}{c}{Syndrome (Depth 3)} & \multicolumn{2}{c}{Syndrome (IP)} & \multicolumn{2}{c}{Greedy Ones ($H_{\text{sum}}$)} & \multicolumn{2}{c}{Greedy Log ($h_{\text{log}}$)} \\
\cmidrule(lr){4-5} \cmidrule(lr){6-7} \cmidrule(lr){8-9} \cmidrule(lr){10-11} \cmidrule(lr){12-13} \cmidrule(lr){14-15} \cmidrule(lr){16-17}
& & & Best choice & Only Choice & Best choice & Only Choice& Best choice  & Only Choice & Best choice & Only Choice & Best choice  & Only Choice & Best choice & Only Choice & Best choice & Only Choice \\
Adder\_8& $ 24$ & 60 & 12 (20\%) & 0 (0)& 13 (22\%) & 0 (0\%)& 42 (70\%) & 0 (0\%)& 42 (70\%) & 0 (0\%)& 42 (70\%) & 0 (0\%)& 60 (100\%) & 2 (3\%)& 58 (97\%) & 0 (0\%)\\
barenco\_tof\_10& $ 19$ & 99 & 29 (29\%) & 0 (0)& 33 (33\%) & 0 (0\%)& 83 (84\%) & 0 (0\%)& 83 (84\%) & 0 (0\%)& 83 (84\%) & 0 (0\%)& 99 (100\%) & 0 (0\%)& 99 (100\%) & 0 (0\%)\\
barenco\_tof\_3& $ 5$ & 14 & 8 (57\%) & 0 (0)& 8 (57\%) & 0 (0\%)& 12 (86\%) & 0 (0\%)& 12 (86\%) & 0 (0\%)& 12 (86\%) & 0 (0\%)& 13 (93\%) & 0 (0\%)& 13 (93\%) & 0 (0\%)\\
barenco\_tof\_4& $ 7$ & 27 & 12 (44\%) & 0 (0)& 12 (44\%) & 0 (0\%)& 24 (89\%) & 0 (0\%)& 24 (89\%) & 0 (0\%)& 24 (89\%) & 0 (0\%)& 26 (96\%) & 0 (0\%)& 26 (96\%) & 0 (0\%)\\
barenco\_tof\_5& $ 9$ & 39 & 16 (41\%) & 0 (0)& 16 (41\%) & 0 (0\%)& 32 (82\%) & 0 (0\%)& 32 (82\%) & 0 (0\%)& 32 (82\%) & 0 (0\%)& 38 (97\%) & 0 (0\%)& 38 (97\%) & 0 (0\%)\\
csla\_mux\_3& $ 15$ & 15 & 0 (0\%) & 0 (0)& 0 (0\%) & 0 (0\%)& 8 (53\%) & 0 (0\%)& 8 (53\%) & 0 (0\%)& 8 (53\%) & 0 (0\%)& 15 (100\%) & 2 (13\%)& 10 (67\%) & 0 (0\%)\\
csum\_mux\_9& $ 30$ & 14 & 1 (7\%) & 0 (0)& 1 (7\%) & 0 (0\%)& 6 (43\%) & 0 (0\%)& 6 (43\%) & 0 (0\%)& 6 (43\%) & 0 (0\%)& 13 (93\%) & 4 (29\%)& 9 (64\%) & 0 (0\%)\\
cycle\_17\_3& $ 35$ & 1436 & 219 (15\%) & 0 (0)& 330 (23\%) & 0 (0\%)& 1094 (76\%) & 0 (0\%)& 1094 (76\%) & 0 (0\%)& 1094 (76\%) & 0 (0\%)& 1429 (100\%) & 23 (2\%)& 1412 (98\%) & 6 (0\%)\\
GF($2^{10}$)\_mult& $ 30$ & 20 & 0 (0\%) & 0 (0)& 0 (0\%) & 0 (0\%)& 1 (5\%) & 0 (0\%)& 1 (5\%) & 0 (0\%)& 1 (5\%) & 0 (0\%)& 16 (80\%) & 13 (65\%)& 6 (30\%) & 3 (15\%)\\
GF($2^{16}$)\_mult& $ 48$ & 28 & 0 (0\%) & 0 (0)& 0 (0\%) & 0 (0\%)& 2 (7\%) & 0 (0\%)& 2 (7\%) & 0 (0\%)& 2 (7\%) & 0 (0\%)& 18 (64\%) & 15 (54\%)& 11 (39\%) & 8 (29\%)\\
GF($2^{32}$)\_mult& $ 96$ & 51 & 0 (0\%) & 0 (0)& 0 (0\%) & 0 (0\%)& 2 (4\%) & 0 (0\%)& 2 (4\%) & 0 (0\%)& 2 (4\%) & 0 (0\%)& 30 (59\%) & 28 (55\%)& 21 (41\%) & 19 (37\%)\\
GF($2^4$)\_mult& $ 12$ & 10 & 0 (0\%) & 0 (0)& 0 (0\%) & 0 (0\%)& 3 (30\%) & 0 (0\%)& 3 (30\%) & 0 (0\%)& 3 (30\%) & 0 (0\%)& 9 (90\%) & 3 (30\%)& 6 (60\%) & 0 (0\%)\\
GF($2^5$)\_mult& $ 15$ & 13 & 1 (8\%) & 0 (0)& 0 (0\%) & 0 (0\%)& 3 (23\%) & 0 (0\%)& 3 (23\%) & 0 (0\%)& 3 (23\%) & 0 (0\%)& 10 (77\%) & 6 (46\%)& 6 (46\%) & 2 (15\%)\\
GF($2^6$)\_mult& $ 18$ & 13 & 0 (0\%) & 0 (0)& 0 (0\%) & 0 (0\%)& 2 (15\%) & 0 (0\%)& 2 (15\%) & 0 (0\%)& 2 (15\%) & 0 (0\%)& 9 (69\%) & 7 (54\%)& 5 (38\%) & 3 (23\%)\\
GF($2^7$)\_mult& $ 21$ & 16 & 0 (0\%) & 0 (0)& 0 (0\%) & 0 (0\%)& 3 (19\%) & 0 (0\%)& 3 (19\%) & 0 (0\%)& 3 (19\%) & 0 (0\%)& 12 (75\%) & 7 (44\%)& 8 (50\%) & 3 (19\%)\\
GF($2^8$)\_mult& $ 24$ & 17 & 0 (0\%) & 0 (0)& 0 (0\%) & 0 (0\%)& 2 (12\%) & 0 (0\%)& 2 (12\%) & 0 (0\%)& 2 (12\%) & 0 (0\%)& 15 (88\%) & 10 (59\%)& 6 (35\%) & 1 (6\%)\\
GF($2^9$)\_mult& $ 27$ & 19 & 0 (0\%) & 0 (0)& 0 (0\%) & 0 (0\%)& 4 (21\%) & 0 (0\%)& 4 (21\%) & 0 (0\%)& 4 (21\%) & 0 (0\%)& 14 (74\%) & 10 (53\%)& 8 (42\%) & 4 (21\%)\\
grover\_5& $ 9$ & 123 & 39 (32\%) & 0 (0)& 39 (32\%) & 0 (0\%)& 110 (89\%) & 0 (0\%)& 110 (89\%) & 0 (0\%)& 110 (89\%) & 0 (0\%)& 122 (99\%) & 0 (0\%)& 121 (98\%) & 0 (0\%)\\
ham15-high& $ 20$ & 852 & 220 (26\%) & 0 (0)& 256 (30\%) & 0 (0\%)& 698 (82\%) & 0 (0\%)& 698 (82\%) & 0 (0\%)& 698 (82\%) & 0 (0\%)& 852 (100\%) & 3 (0\%)& 849 (100\%) & 0 (0\%)\\
ham15-low& $ 17$ & 69 & 21 (30\%) & 0 (0)& 18 (26\%) & 0 (0\%)& 57 (83\%) & 0 (0\%)& 57 (83\%) & 0 (0\%)& 57 (83\%) & 0 (0\%)& 69 (100\%) & 2 (3\%)& 66 (96\%) & 0 (0\%)\\
ham15-med& $ 17$ & 189 & 56 (30\%) & 0 (0)& 63 (33\%) & 0 (0\%)& 167 (88\%) & 0 (0\%)& 167 (88\%) & 0 (0\%)& 167 (88\%) & 0 (0\%)& 189 (100\%) & 1 (1\%)& 188 (99\%) & 0 (0\%)\\
hwb6& $ 7$ & 48 & 11 (23\%) & 0 (0)& 12 (25\%) & 0 (0\%)& 40 (83\%) & 0 (0\%)& 40 (83\%) & 0 (0\%)& 40 (83\%) & 0 (0\%)& 47 (98\%) & 0 (0\%)& 47 (98\%) & 0 (0\%)\\
hwb8& $ 12$ & 2130 & 270 (13\%) & 0 (0)& 343 (16\%) & 0 (0\%)& 1248 (59\%) & 0 (0\%)& 1245 (58\%) & 0 (0\%)& 1247 (59\%) & 0 (0\%)& 2115 (99\%) & 159 (7\%)& 1938 (91\%) & 10 (0\%)\\
mod5\_4& $ 5$ & 14 & 8 (57\%) & 0 (0)& 8 (57\%) & 0 (0\%)& 13 (93\%) & 0 (0\%)& 13 (93\%) & 0 (0\%)& 13 (93\%) & 0 (0\%)& 14 (100\%) & 0 (0\%)& 14 (100\%) & 0 (0\%)\\
mod\_adder\_1024& $ 28$ & 716 & 173 (24\%) & 0 (0)& 214 (30\%) & 0 (0\%)& 564 (79\%) & 0 (0\%)& 564 (79\%) & 0 (0\%)& 564 (79\%) & 0 (0\%)& 716 (100\%) & 28 (4\%)& 688 (96\%) & 0 (0\%)\\
mod\_adder\_1048576& $ 58$ & 5615 & 857 (15\%) & 0 (0)& 1335 (24\%) & 0 (0\%)& 4275 (76\%) & 0 (0\%)& 4276 (76\%) & 0 (0\%)& 4276 (76\%) & 0 (0\%)& 5593 (100\%) & 114 (2\%)& 5500 (98\%) & 21 (0\%)\\
mod\_mult\_55& $ 9$ & 14 & 4 (29\%) & 0 (0)& 4 (29\%) & 0 (0\%)& 9 (64\%) & 0 (0\%)& 9 (64\%) & 0 (0\%)& 9 (64\%) & 0 (0\%)& 14 (100\%) & 0 (0\%)& 14 (100\%) & 0 (0\%)\\
mod\_red\_21& $ 11$ & 46 & 19 (41\%) & 0 (0)& 19 (41\%) & 0 (0\%)& 40 (87\%) & 0 (0\%)& 40 (87\%) & 0 (0\%)& 40 (87\%) & 0 (0\%)& 45 (98\%) & 0 (0\%)& 44 (96\%) & 0 (0\%)\\
qcla\_adder\_10& $ 36$ & 24 & 3 (12\%) & 0 (0)& 4 (17\%) & 0 (0\%)& 17 (71\%) & 0 (0\%)& 17 (71\%) & 0 (0\%)& 17 (71\%) & 0 (0\%)& 23 (96\%) & 2 (8\%)& 20 (83\%) & 0 (0\%)\\
qcla\_com\_7& $ 24$ & 26 & 4 (15\%) & 0 (0)& 4 (15\%) & 0 (0\%)& 15 (58\%) & 0 (0\%)& 15 (58\%) & 0 (0\%)& 15 (58\%) & 0 (0\%)& 26 (100\%) & 0 (0\%)& 26 (100\%) & 0 (0\%)\\
qcla\_mod\_7& $ 26$ & 52 & 5 (10\%) & 0 (0)& 5 (10\%) & 0 (0\%)& 35 (67\%) & 0 (0\%)& 35 (67\%) & 0 (0\%)& 35 (67\%) & 0 (0\%)& 51 (98\%) & 2 (4\%)& 48 (92\%) & 0 (0\%)\\
qft\_4& $ 5$ & 29 & 9 (31\%) & 0 (0)& 12 (41\%) & 0 (0\%)& 28 (97\%) & 0 (0\%)& 28 (97\%) & 0 (0\%)& 28 (97\%) & 0 (0\%)& 29 (100\%) & 0 (0\%)& 29 (100\%) & 0 (0\%)\\
rc\_adder\_6& $ 14$ & 49 & 33 (67\%) & 0 (0)& 34 (69\%) & 0 (0\%)& 43 (88\%) & 0 (0\%)& 43 (88\%) & 0 (0\%)& 43 (88\%) & 0 (0\%)& 48 (98\%) & 0 (0\%)& 48 (98\%) & 0 (0\%)\\
tof\_10& $ 19$ & 52 & 21 (40\%) & 0 (0)& 21 (40\%) & 0 (0\%)& 48 (92\%) & 0 (0\%)& 48 (92\%) & 0 (0\%)& 48 (92\%) & 0 (0\%)& 51 (98\%) & 0 (0\%)& 50 (96\%) & 0 (0\%)\\
tof\_3& $ 5$ & 12 & 7 (58\%) & 0 (0)& 8 (67\%) & 0 (0\%)& 11 (92\%) & 0 (0\%)& 11 (92\%) & 0 (0\%)& 11 (92\%) & 0 (0\%)& 11 (92\%) & 0 (0\%)& 11 (92\%) & 0 (0\%)\\
tof\_4& $ 7$ & 17 & 10 (59\%) & 0 (0)& 11 (65\%) & 0 (0\%)& 16 (94\%) & 0 (0\%)& 16 (94\%) & 0 (0\%)& 16 (94\%) & 0 (0\%)& 16 (94\%) & 0 (0\%)& 16 (94\%) & 0 (0\%)\\
tof\_5& $ 9$ & 22 & 11 (50\%) & 0 (0)& 11 (50\%) & 0 (0\%)& 18 (82\%) & 0 (0\%)& 18 (82\%) & 0 (0\%)& 18 (82\%) & 0 (0\%)& 21 (95\%) & 0 (0\%)& 21 (95\%) & 0 (0\%)\\
vbe\_adder\_3& $ 10$ & 19 & 4 (21\%) & 0 (0)& 4 (21\%) & 0 (0\%)& 18 (95\%) & 0 (0\%)& 18 (95\%) & 0 (0\%)& 18 (95\%) & 0 (0\%)& 18 (95\%) & 0 (0\%)& 18 (95\%) & 0 (0\%)\\
\bottomrule
\end{tabular}
}
\label{bench::frequency}
\end{sidewaystable*}

\section{Conclusion} \label{cnot_size::conclusion} We presented the
simple algorithm GreedyGE for the synthesis of linear reversible
circuits. We improved state-of-the-art algorithms by adding a greedy
feature in the Gaussian elimination algorithm while keeping an overall
structure in the synthesis process for triangular operators. We
combined this method with a practical LU decomposition that improves
the results for input circuits of small sizes. Overall our method is
fast and provides the best results when $n$ is sufficiently large
($n > 150$). For operators acting on a small number of qubits
($n < 30$) or when we expect the operators to be synthesized with a
small circuit, then purely greedy methods have shown to give quasi
optimal results. We also managed to significantly reduce the CNOT
count and, surprisingly, the depth of well-known reversible functions
while keeping the T-count and the T-depth as low as possible, as given
by other optimization algorithms \cite{DBLP:journals/tcad/AmyMM14}.

Theoretically, GreedyGE worst case complexity is guaranteed to be at
most asymptotically equal to $n^2/\log_2(n)$, which is still a factor
of 2 larger than the theoretical lower bound. We will investigate in
future work if this theoretical lower bound can be reached. Another
main issue will be to extend this algorithm to the case where the
qubits connectivity follows a restricted topology. It would also be
interesting to look at the performance of GreedyGE for the synthesis
of CNOT+T circuits. It is known that the synthesis of a CNOT+T circuit
can be performed by manipulating a rectangular parity table via row
operations \cite{amy2018controlled}. Although the goal is not exactly
to reduce the parity table to an identity operator, it involves to
reduce the Hamming weight of the columns to $1$ and to remove them
from the table until the table is empty. Given that GreedyGE, at each
step, reduces the Hamming weight of a column to $1$, it can be used to
CNOT+T circuits synthesis as well.

\section*{Acknowledgment}
This work was supported in part by the French National Research Agency
(ANR) under the research project SoftQPRO ANR-17-CE25-0009-02, and by
the DGE of the French Ministry of Industry under the research project
PIA-GDN/QuantEx P163746-484124.

\newpage

\bibliographystyle{acm}
\bibliography{Biblio.bib}

\end{document}